\documentclass[11pt]{article}%
\usepackage[margin=1in]{geometry}
\usepackage{amsmath}
\usepackage{amsfonts}
\usepackage{amssymb}
\usepackage{graphicx}
\usepackage{subcaption}
\usepackage{float}
\usepackage{epstopdf}
\usepackage{authordate1-4}%
\usepackage{hyperref}
\usepackage{authblk}

\begin{document}

\title{Modal interactions in strongly nonlinear coupled oscillators with closely spaced modes and non-classical damping distribution}

\author[1]{Csanád Á. Hubay\thanks{Corresponding author: hubay.csanad@gmail.com}}
\author[1]{Tamás Kalmár-Nagy}
\author[2]{Lawrence A. Bergman}
\author[3]{Alexander F. Vakakis}

\affil[1]{Department of Fluid Mechanics, Faculty of Mechanical Engineering, Budapest University of Technology and Economics, Műegyetem rkp. 3., Budapest H-1111, Hungary}
\affil[2]{Department of Aerospace Engineering, The Grainger College of Engineering, University of Illinois Urbana-Champaign, 104 S Wright St, Urbana, IL 61801, USA}
\affil[3]{Department of Mechanical Science and Engineering, The Grainger College of Engineering, University of Illinois Urbana-Champaign, 1206 W Green St, Urbana, IL 61801, USA}

\date{}
\maketitle

\begin{abstract}

Linear and nonlinear mechanical structures with closely spaced modes and non-classical (non-proportional) damping distribution are susceptible to modal interactions with energy exchanges between modes. In this work we analyze these interactions in a two-degree-of-freedom (2 DoF) coupled oscillator subject to impulse excitation, utilizing action-angle variables and the method of multiple scales. To show the basic steps of the methodology we start from the linear system, then proceed to analyze modal interactions in the system with strongly nonlinear cubic stiffness nonlinearity. In contrast to the linear case, once the dynamics of the nonlinear system is captured in 1:1 resonance, strongly nonlinear modal interactions occur depending on energy. These regimes as well as attraction to capture, 1:1 capture and escape from capture are analyzed in detail. Our analytical study determines the ranges of parameter values for which nonlinear modal interactions due to resonance capture occur. The analytical predictions are fully verified by direct numerical simulations of the governing equations of motion. Possible extensions of the present work to systems of oscillators with more DoF and other types of dissipative elements, e.g., friction or hysteresis are discussed.
\bigskip


\end{abstract}

\section{Introduction}

In engineering practice many mechanical structures, for instance, bladed disk assemblies in turbomachinery \cite{ewins2000modal}, can be modelled as linear or nonlinear multi-degree-of-freedom (mDoF) coupled oscillators with families of closely spaced modes. This type of modes commonly arises from structural symmetries or particular combinations of mass and stiffness distributions. For linear, classically damped coupled oscillators  (often referred to as systems with proportional damping distribution) these modes are uncoupled \cite{caughey1960classical,caughey1965classical,phani2003necessary}, and each mode evolves within a two-dimensional invariant subspace of the system's mode space. In contrast, a non-classical damping distribution (non-proportional damping) couples the resulting complex vibration modes \cite{ewins2000modal,krack2016motion} enabling \emph{modal interactions} and energy exchanges between modes. From a geometrical point of view, the two-dimensional invariant manifolds in the mode space of  the uncoupled vibration modes that exist for classical damping, combine into higher-dimensional invariant manifolds for non-classical damping distribution. A need therefore exists for better interpretation of parameter-dependent modal interactions and modal energy exchanges in mDoF coupled oscillators; in the linear case this was recently addressed in \cite{pinto2026understanding} for a two-dimensional system of coupled oscillators with closely spaced modes and non-classical damping distribution.

The aim of the present study is to provide an alternative analytical framework for addressing parameter-dependent modal interactions and modal energy exchanges in mDoF nonlinear, coupled oscillators, based on action–angle variables \cite{percival1982introduction} and multiple scales analysis \cite{nayfeh1984nonlinear}. We apply this method to analyze modal interactions in an impulsively excited two-degree-of-freedom (2 DoF) system of coupled oscillators with closely spaced modes and non-classical damping, considering first the linear case and subsequently the strongly nonlinear case. In the linear case we recover the results of \cite{pinto2026understanding}, and then extend the study to the system with strong cubic stiffness nonlinearity. We investigate the ranges of the system parameters as well as the impulse intensity (input energy) for which nonlinear modal interactions occur due to a 1:1 transient resonance capture (TRC). In addition, we determine the existence criterion for a 1:1 resonant manifold in the transient dynamics, and formulate necessary and sufficient conditions for realization of recurring nonlinear beat phenomena. When this criterion is met, the responses are characterized by a single effective dissipation rate and two distinct, closely spaced, fast oscillation frequencies. The superposition of these two fast-frequency oscillations produces a slow amplitude modulation (beating phenomenon). 

\section{Mechanical model}
\label{sec:model}

The mechanical model of interest consists of two identical masses $m_{1}$ and $m_2$ (i.e., $m_1=m_2$) with grounding stiffnesses $k_{1}$ or $k_{\mathrm{nl}}$, and $ k_2$, respectively, and a weak coupling stiffness $\varepsilon k_c$. The masses are also attached to grounding viscous dampers with damping coefficients $\varepsilon c_{1}$ and $\varepsilon c_{2}$. The coupling $\varepsilon k_c$ and damping coefficients $\varepsilon c_{1}, \varepsilon c_{2}$ are of the same (small) order of magnitude, and characterized by the \textquoteleft small\textquoteright~parameter $\varepsilon$. Then setting $\frac{c_1}{m_1}=\lambda_1$, $\frac{c_2}{m_2}=\lambda_2$, $\frac{k_c}{m_1}=\frac{k_c}{m_2}=\beta$, $\frac{k_1}{m_1}=\beta_1$, $\frac{k_2}{m_2}=\beta_2=1$, and $\frac{k_{\mathrm{nl}}}{m_1}=\alpha$ yields the equation of motion,
\begin{equation}
    \begin{aligned}
      &\ddot x_1(t)+\varepsilon \lambda_1\dot x_1(t)+\beta_1x_1(t)+\varepsilon \beta(x_1(t)- x_2(t))+\alpha x_1^3(t)=0\\
       &\ddot x_2(t)+\varepsilon \lambda_2\dot x_2(t)+\beta_2x_2(t)+\varepsilon \beta(x_2(t)- x_1(t))=0 
    \end{aligned} \label{eq:x1x2base}
\end{equation}
with the initial conditions,
\begin{equation}
        x_1 (0)=x_2 (0)=0,\quad \dot{x}_1 (0)=v_1,\quad \dot{x}_2 (0)=v_2\label{eq:ini}
\end{equation}
where $v_1\geq0$ and $v_2\geq0$ are the \emph{impulse intensities}. 

In this work, two particular cases of system~\eqref{eq:x1x2base} are investigated. First, we consider the \emph{linear case}, for which,
\begin{equation}
\alpha=0,\qquad \beta_1=1
\end{equation}
meaning that the uncoupled oscillators have identical natural angular frequencies so that 1:1 resonance can occur in system~\eqref{eq:x1x2base}. Second, we consider the \emph{strongly nonlinear case},
\begin{equation}
\alpha>0,\qquad \beta_1=0
\end{equation}
i.e., mass $m_1$ is grounded solely through a nonlinear stiffness. This case allows us to study strongly nonlinear modal interactions in the system with a non-linearizable cubic stiffness nonlinearity.

The system has non-classical damping distribution that represents a more general setting for system~\eqref{eq:x1x2base} and it introduces modal coupling. Our analysis is restricted to this case. For the linear case, the symmetry, $m_1=m_2$ and $\beta_1=\beta_2$, allows us, without loss of generality, to consider that $\lambda_1>\lambda_2$. In contrast, for the nonlinear system, the two cases $\lambda_1>\lambda_2$ and $\lambda_1<\lambda_2$ have to be treated separately.

\section{Analysis of modal interactions in the linear system}
\label{sec:lin}

Considering the 2 DoF linear, coupled system with $\alpha=0$ and $\beta_1=1$, the equations of motion are:
\begin{equation}
    \begin{aligned}
        \ddot x_1(t)+\varepsilon\lambda_1\dot x_1(t)+x_1(t)+\varepsilon\beta(x_1(t)- x_2(t))&=0\\
        \ddot x_2(t)+\varepsilon\lambda_2\dot x_2(t)+x_2(t)+\varepsilon\beta(x_2(t)- x_1(t))&=0\\
        x_1 (0)=x_2 (0)=0,\quad \dot{x}_1 (0)=v_1,\quad \dot{x}_2 (0)&=v_2
    \end{aligned}\label{eq:x1x2eom}
\end{equation}
In the following, we investigate the free-decaying response of system~\eqref{eq:x1x2eom} subject to a unit impulse excitation applied to either of the two masses. Specifically, we consider the two cases of impulse intensities, $v_1=\sqrt{2},v_2=0$ and $v_1=0,v_2=\sqrt{2}$. 

We introduce at this point the following action-angle transformation utilizing the actions and angles of the underlying uncoupled, undamped and unforced system,  \cite{percival1982introduction},
\begin{equation}
    \begin{aligned}
    x_1(t)&=\sqrt{2 I_1(t)}\sin\theta_1(t),\quad \dot{x}_1(t)=\sqrt{2I_1(t)}\cos\theta_1(t)\\
    x_2(t)&=\sqrt{2I_2(t)}\sin\theta_2(t),\quad \dot{x}_2(t)=\sqrt{2I_2(t)}\cos\theta_2(t)\label{eq:x1x2Def}
    \end{aligned}
\end{equation}
in terms of which the two second-order equations of motion~\eqref{eq:x1x2eom} are transformed into the following set of four first-order equations,
\begin{equation}
    \begin{aligned}
    \dot{I}_1(t)&=\varepsilon\left(-2\lambda_1 I_1(t)\cos^2\theta_1(t)-\beta I_1(t) \sin 2\theta_1(t)+2\beta\sqrt{I_1(t)I_2(t)}\cos\theta_1(t)\sin\theta_2(t)\right)\\
   \dot{I}_2(t)&= \varepsilon\left(-2\lambda_2 I_2(t)\cos^2\theta_2(t)-\beta I_2(t) \sin 2\theta_2(t)+2\beta\sqrt{I_1(t)I_2(t)}\cos\theta_2(t)\sin\theta_1(t)\right)\\
    \dot{\theta}_1(t)&=1+\varepsilon\left(\frac{\lambda_1}{2}  \sin2 \theta_1(t)+\beta\sin^2\theta_1(t)-\beta \sqrt{ \frac{I_2(t)}{I_1(t)}} \sin \theta_1 (t)\sin \theta_2 (t)\right)\\
\dot{\theta}_2(t)&=1+\varepsilon\left(\frac{\lambda_2}{2} \sin2\theta_2(t)+\beta\sin^2\theta_2(t)-\beta\sqrt{\frac{I_1(t)}{I_2(t)}}\sin\theta_1(t)\sin\theta_2(t)\right) \label{eq:x1x2angles}  
    \end{aligned}
\end{equation}
with initial conditions:
\begin{gather}
    I_1(0)=\frac{v_1^2}{2},\quad I_2(0)=\frac{v_2^2}{2}\label{eq:actini}\\
   \theta_1(0)=\left\{\begin{array}[c]{c}
        0,\quad v_1\neq0, v_2=0\\
        -\frac{\pi}{2},\quad v_1=0, v_2\neq0\\
    \end{array}\right.,\quad \theta_2(0)=\left\{\begin{array}[c]{c}
        -\frac{\pi}{2},\quad v_1\neq0, v_2=0\\
        0,\quad v_1=0, v_2\neq0\\
    \end{array}\right. \label{eq:actini2}
\end{gather}
It follows that for weak coupling $\varepsilon\beta$ and damping $\varepsilon\lambda_1,\varepsilon\lambda_2$, system~\eqref{eq:x1x2eom} exhibits two instantaneous oscillation frequencies $\dot{\theta}_1(t)$ and $\dot{\theta}_2(t)$ close to the respective uncoupled angular frequencies, which are equal to unity. For small values of the parameter $\varepsilon$, a 1:1 internal resonance can be realized in the  dynamics of system~\eqref{eq:x1x2angles}, with the oscillation possessing a \emph{fast} frequency close to unity. As the coupling $\varepsilon\beta$ is increased the difference between the two instantaneous oscillation frequencies $\dot{\theta}_1(t)$ and $\dot{\theta}_2(t)$ also increases, but remains small compared to the fast frequency, so the condition of 1:1 internal resonance still holds. Consequently, the relative phase between $\theta_1(t)$ and $\theta_2(t)$ remains small and evolves on a \emph{slow} timescale introducing a \emph{slow phase} (which is characteristic of 1:1 internal resonance), yielding slowly modulated envelopes in the corresponding oscillations of the system.

In the following, we restrict our analysis to sufficiently small values of the parameter $\varepsilon$, so that the dynamics of system~\eqref{eq:x1x2angles} remains in the regime of 1:1 internal resonance, with the dynamics possessing slow and fast time scales, as described above. To this end, we define the slow angle, 
\begin{equation}
    \psi(t)=\theta_1(t)-\theta_2(t)\label{eq:psi}
\end{equation}
which, upon substitution into Eq.~\eqref{eq:x1x2angles} yields the following set of equations that are valid in the neighborhood of 1:1 internal resonance:
\begin{equation}
    \begin{aligned}
    \dot{I}_1(t)&=\varepsilon\Big(-2\lambda_1 I_1(t)\cos^2\left(\theta_2(t)+\psi(t)\right)-\\ & -\beta I_1(t) \sin\left(2\theta_2(t)+2\psi(t)\right)+\beta\sqrt{I_1(t)I_2(t)}\big(\sin(2\theta_2(t)+\psi(t))-\sin\psi(t)\big)\Big)\\
   \dot{I}_2(t)&= \varepsilon\Big(-2\lambda_2 I_2(t)\cos^2\theta_2(t)-  \\ &-\beta I_2(t) \sin 2\theta_2(t)+\beta\sqrt{I_1(t)I_2(t)}\big(\sin(2\theta_2(t)+\psi(t))+\sin\psi(t)\big)\Big)\\
   \dot\psi(t)&=\varepsilon\Bigg(\frac{\lambda_1}{2}\sin\big(2\theta_2(t)+2\psi(t)\big)-\frac{\lambda_2}{2}\sin2\theta_2(t)+\beta\left(\sin^2\left(\theta_2(t)+\psi(t)\right)-\sin^2\theta_2(t)\right)-\\
   &-\frac{\beta}{2}\left(\sqrt{\frac{I_2(t)}{I_1(t)}}-\sqrt{\frac{I_1(t)}{I_2(t)}}\right)\big(\cos\psi(t)-\cos(2\theta_2(t)+\psi(t))\big)\Bigg)\label{eq:x1x24eqs}
    \end{aligned}
\end{equation}
These slow modulation equations govern the slow evolutions of the actions $I_1(t),I_2(t)$ and phase ${\psi}(t)$, and capture all the important dynamics of 1:1 resonance in the system of coupled oscillators. At this point we note that, from Eq.~\eqref{eq:x1x2angles} the second (fast) angle variable is governed by,
\begin{equation}
    \dot{\theta}_2(t)=1+\varepsilon\left(\frac{\lambda_2}{2}\sin2\theta_2(t)+\beta\sin^2\theta_2(t)-\frac{\beta}{2}\sqrt{\frac{I_1(t)}{I_2(t)}}\big(\cos\psi(t)-\cos\left(2\theta_2(t)+\psi(t)\right)\big)\right)
\end{equation}
so it is a \textquotedblleft time-like\textquotedblright~(monotonically increasing) variable. Indeed, direct integration yields,
\begin{equation}
    t=\theta_2(t)+\mathcal{O}(\varepsilon) \label{eq:x1x2time}
\end{equation}
so we can replace the time variable $t$ with $\theta_2$ as the independent variable in the slow modulation equations~\eqref{eq:x1x24eqs}. Following that, we transform the system into autonomous form and apply the method of multiple scales, by introducing the fast and slow scales:
\begin{equation}
    T_0=\theta_2,\qquad T_1=\varepsilon \theta_2
\end{equation}
Then, the asymptotic approximations to the solutions of Eq.~\eqref{eq:x1x24eqs} are sought in the form,
\begin{equation}
    \begin{aligned}
        I_1(\theta_2)=I_1^0(T_0,T_1)+\varepsilon I_1^1(T_0,T_1)+\mathcal{O}(\varepsilon^2)\\
        I_2(\theta_2)=I_2^0(T_0,T_1)+\varepsilon I_2^1(T_0,T_1)+\mathcal{O}(\varepsilon^2)\\
        \psi(\theta_2)=\psi^0(T_0,T_1)+\varepsilon \psi^1(T_0,T_1)+\mathcal{O}(\varepsilon^2) \label{eq:ansatz}
    \end{aligned}
\end{equation}
which, after substituting into Eq.~\eqref{eq:x1x24eqs} yield the following leading approximations at order $\varepsilon^0$:
\begin{equation}
    \begin{aligned}
       \frac{\partial I_1^0}{\partial T_0} &=0 \qquad \rightarrow \qquad  I_1^0(T_0,T_1)=A_0(T_1)\\
   \frac{\partial I_2^0}{\partial T_0} &=0\qquad  \rightarrow \qquad I_2^0(T_0,T_1)=B_0(T_1)\\
   \frac{\partial \psi^0}{\partial T_0} &=0\qquad  \rightarrow \qquad \psi^0(T_0,T_1)=C_0(T_1) \label{eq:x1x2zeroth}
    \end{aligned}
\end{equation}
Consequently, the $\varepsilon^1$ approximations are governed by:
\begin{equation}
    \begin{aligned}
       \frac{\partial I_1^1}{\partial T_0} &=-A_0^\prime(T_1) -2\lambda_1 A_0(T_1)\cos^2\left(T_0+C_0(T_1)\right)-\\ & -\beta A_0(t) \sin\left(2T_0+2C_0(T_1)\right)+\beta\sqrt{A_0(T_1)B_0(T_1)}\big(\sin(2T_0+C_0(T_1))-\sin C_0(T_1)\big)\\
   \frac{\partial I_2^1}{\partial T_0} &=-B_0^\prime(T_1) -2\lambda_2 B_0(T_1)\cos^2T_0- \\ &-\beta B_0(t) \sin 2T_0+\beta\sqrt{A_0(T_1)B_0(T_1)}\big(\sin(2T_0+C_0(T_1))+\sin C_0(T_1)\big)\\
   \frac{\partial \psi^1}{\partial T_0} &=-C_0^\prime(T_1)+\frac{\lambda_1}2\sin\big(2T_0+2C_0(T_1)\big)-\frac{\lambda_2}2\sin2T_0+\beta\left(\sin^2\left(T_0+C_0(T_1)\right)-\sin^2T_0\right)-\\
   &\qquad-\frac{\beta}{2}\left(\sqrt{\frac{B_0(T_1)}{A_0(T_1)}}-\sqrt{\frac{A_0(T_1)}{B_0(T_1)}}\right)\big(\cos C_0(T_1)-\cos(2T_0+C_0(T_1))\big)\label{eq:firstorder}
    \end{aligned}
\end{equation}
We impose the solvability conditions to eliminate the secular terms from the solution of \eqref{eq:firstorder}, which yields the following slow-flow equations in the slow time scale $T_1$:
\begin{equation}
    \begin{aligned}
       A_0^\prime(T_1)&=-\lambda_1A_0(T_1)-\beta\sqrt{A_0(T_1)B_0(T_1)}\sin C_0(T_1) \\
  B_0^\prime(T_1)&=-\lambda_2B_0(T_1)+\beta\sqrt{A_0(T_1)B_0(T_1)}\sin C_0(T_1)\\
   C_0^\prime(T_1)&=-\frac{\beta}{2}\left(\sqrt{\frac{B_0(T_1)}{A_0(T_1)}}-\sqrt{\frac{A_0(T_1)}{B_0(T_1)}}\right)\cos C_0(T_1) 
    \end{aligned}\label{eq:x1x2solvability}
\end{equation}
Eq.~\eqref{eq:x1x2solvability} governs the leading order analytical approximations to the slow-varying envelopes of the oscillations of the two masses of the system coupled through the stiffness $\varepsilon\beta$. The initial conditions for the slow-flow equations are (cf. Eqs.~\eqref{eq:actini} and~\eqref{eq:actini2})
\begin{equation}
    A_0(0)=\frac{v_1^2}{2},\quad B_0(0)=\frac{v_2^2}{2},\quad C_0(0)=\left\{\begin{array}[c]{c}
        \frac{\pi}{2},\quad v_1\neq0, v_2=0\\
        -\frac{\pi}{2},\quad v_1=0, v_2\neq0\\
    \end{array}\right. \label{eq:slowini}
\end{equation} 
Eq.~\eqref{eq:x1x2solvability} can be solved by introducing the (dependent) variables
\begin{equation}
    \begin{aligned}
         X(T_1)=A_0(T_1)+B_0(T_1),&\quad Y(T_1)=A_0(T_1)-B_0(T_1)\\
         \mathrm{Re}(Z(T_1))=\sqrt{A_0(T_1)B_0(T_1)}\cos(C_0(T_1)),&\quad \mathrm{Im}(Z(T_1))=\sqrt{A_0(T_1)B_0(T_1)}\sin(C_0(T_1))
    \end{aligned}
    \label{eq:x1x2trans}
\end{equation}
which can be interpreted in terms of the system coordinates $x_1(t)$ and $x_2(t)$, and also, in terms of the modal coordinates
\begin{equation}
s(t)=\frac{x_1(t)+x_2(t)}{\sqrt{2}}, \quad a(t)=\frac{x_1(t)-x_2(t)}{\sqrt{2}} \label{eq:modDef}
\end{equation}
corresponding to the in-phase and out-of-phase modes of system~\eqref{eq:x1x2eom}, respectively. The energies of the in-phase and out-of-phase modes in terms of variables~\eqref{eq:x1x2trans} are:
\begin{equation}
    E_s(T_1)=\frac{1}{2}X(T_1)+\mathrm{Re}(Z(T_1)),\quad E_a(T_1)=\frac{1}{2}X(T_1)-\mathrm{Re}(Z(T_1))
    \label{eq:modalEnergies}
\end{equation}
Thus, $X(T_1)$ represents the total energy of system~\eqref{eq:x1x2eom}, which can be decomposed either between the DoFs, $X(T_1)=A_0(T_1)+B_0(T_1)$, or between the modes, $X(T_1)=E_s(T_1)+E_a(T_1)$. Similarly, $Y(T_1)$ represents the energy difference between the two DoFs, while $\mathrm{Re}(Z(T_1))=\frac{1}{2}(E_s(T_1)-E_a(T_1))$ represents the energy difference between the in-phase and out-of-phase modes. The variable $\mathrm{Im}(Z(T_1))$ encapsulates the phase interaction between the two DoFs. Then, the slow modulation equations in the new variables are expressed as,
\begin{equation}
    \begin{aligned}
       X^\prime(T_1)&=-\lambda_sX(T_1)-\lambda_aY(T_1) \\
       Y^\prime(T_1)&=-\lambda_aX(T_1)-\lambda_sY(T_1)-2\beta\mathrm{Im}(Z(T_1))\\
        \mathrm{Re}(Z^\prime(T_1))&=-\lambda_s\mathrm{Re}(Z(T_1))\\
        \mathrm{Im}(Z^\prime(T_1))&=-\lambda_s\mathrm{Im}(Z(T_1))+\frac{\beta}{2}Y(T_1) \label{eq:x1x2slowEqs}
    \end{aligned}
\end{equation}
where $\lambda_s$ represents the overall energy decay rate, $\beta$ governs the energy exchanges between the two DoFs of the system, and $\lambda_a$ represents the coupling between modes induced by the non-classical damping distribution and resulting in modal interactions. In particular, the energy $X(T_1)$ and the energy difference $Y(T_1)$ coupled through $\lambda_a$ enable redistribution of (the initial) energy between the two DoFs, whereas for a classical damping distribution ($\lambda_1=\lambda_2$), it holds that $\lambda_a=0$, so the variables $X(T_1)$ and $Y(T_1)$ decouple and each DoF decays independently at the rate $\lambda_s$. The initial conditions of Eq.~\eqref{eq:x1x2slowEqs} are determined from Eq.~\eqref{eq:slowini} based on the variable change~\eqref{eq:x1x2trans} as follows:
\begin{equation}
    X_0\equiv X(0)=\frac{v_1^2+v_2^2}{2},\quad  Y_0\equiv Y(0)=\frac{v_1^2-v_2^2}{2}, \quad Z_0\equiv Z(0)=\left\{\begin{array}[c]{c}
        i\frac{v_1v_2}{2},\quad v_1\neq0, v_2=0\\
        -i\frac{v_1v_2}{2},\quad v_1=0, v_2\neq0\\
    \end{array}\right. \label{eq:finalini}
\end{equation}

Denoting $\nu=\sqrt{\lambda_a^2-\beta^2}$, the exact solutions of the linear system~\eqref{eq:x1x2slowEqs} for the initial conditions~\eqref{eq:finalini} are:
\begin{equation}
    \begin{aligned}
       X(T_1)&=e^{-\lambda_s T_1}\left(\frac{X_0}{\nu^2}\big(\lambda_a^2\cosh\left(\nu T_1\right)-\beta^2\big)-\frac{\lambda_aY_0}{\nu}\sinh\left(\nu T_1\right)\right)\\
       Y(T_1)&=e^{-\lambda_s T_1}\left(Y_0\cosh\left(\nu T_1\right)-\frac{\lambda_a X_0}{\nu}\sinh\left(\nu T_1\right)\right)\\
       Z(T_1)&=e^{-\lambda_s T_1}\left(Z_0+i\frac{X_0}{2\nu^2}\left(\lambda_a\beta -\lambda_a \beta\cosh\left(\nu T_1\right)+\beta\nu\frac{ Y_0}{X_0}\sinh\left(\nu T_1\right)\right)\right)\label{eq:x1x2partSol}
    \end{aligned}
\end{equation}
Then, the leading-order approximation of the energy $E(t)$ of system~\eqref{eq:x1x2eom} can be expressed as:
\begin{equation}
E(t)=X(\varepsilon t)+\mathcal{O}(\varepsilon)
\end{equation} 
So, for the impulse applied to the left mass $m_1$, i.e., for $v_1=\sqrt{2},v_2=0$, the total energy decay is approximated as,
\begin{equation}
    E(t)=e^{-\varepsilon\lambda_s t}\left(\frac{v_1^2}{2\nu^2}\big(\lambda_a^2\cosh\left( \varepsilon \nu t\right)-\beta^2\big)-\frac{\lambda_av_1^2}{2\nu}\sinh\left(\varepsilon \nu t\right)\right)+\mathcal{O}(\varepsilon)\label{eq:E1}
\end{equation} 
while for the impulse applied to the right mass $m_2$, i.e., for initial velocities $v_1=0,v_2=\sqrt{2}$: 
\begin{equation}
    E(t)=e^{-\varepsilon\lambda_s t}\left(\frac{v_2^2}{2\nu^2}\big(\lambda_a^2\cosh\left( \varepsilon \nu t\right)-\beta^2\big)+\frac{\lambda_av_2^2}{2\nu}\sinh\left(\varepsilon \nu t\right)\right)+\mathcal{O}(\varepsilon)\label{eq:E2}
\end{equation} 
From the structure of the solutions~\eqref{eq:x1x2partSol}-\eqref{eq:E2} it follows that in the neighborhood of 1:1 internal resonance system~\eqref{eq:x1x2eom} has two regimes of dynamics with different decay characteristics. For $\beta<\lambda_a$, the parameter $\nu$ is real-valued and the hyperbolic functions in the responses indicate that there exist three distinct exponential decay rates, namely, $
\lambda_s-\nu,
\lambda_s,
\lambda_s+\nu$.
 Furthermore, the total energy decay is dependent on the location of the initial excitation. In contrast, for $\beta>\lambda_a$, $\nu$ is purely imaginary, and the hyperbolic functions transform to bounded trigonometric functions, with the result that $\lambda_s$ is the only exponential decay rate in the total energy decay. The transition between these two distinct dynamical regimes occurs through a bifurcation at $
    \beta_{\mathrm{crit}}=\lambda_a $.

In \cite{pinto2026understanding} this same system was studied using a different asymptotic methodology, and the \emph{bifurcation} was found to occur when, $\gamma=\frac{4\beta}{\lambda_1-\lambda_2}=\frac{2\beta}{\lambda_a}=2$; this condition corresponds exactly to the bifurcation value derived here, $\beta_{\mathrm{crit}}=\lambda_a$. Hence, the previous action-angle based analysis fully recovers the results of that previous work.

\begin{figure}[ptb]
    \begin{subfigure}{0.48\textwidth}
        \centering
        \includegraphics[width=\linewidth]{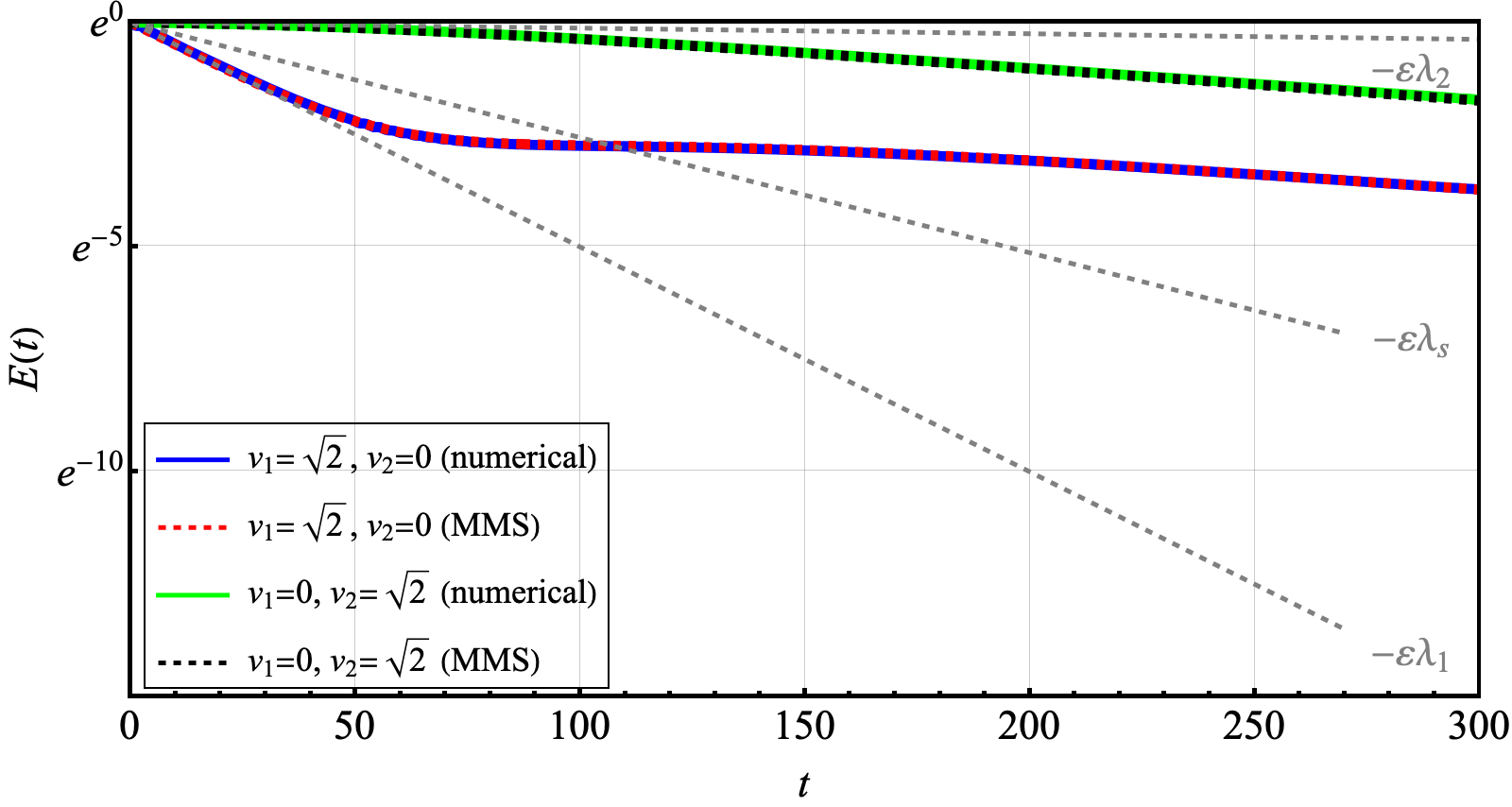}
        \caption{}
    \end{subfigure}
    \hfill
    \begin{subfigure}{0.48\textwidth}
        \centering
        \includegraphics[width=\linewidth]{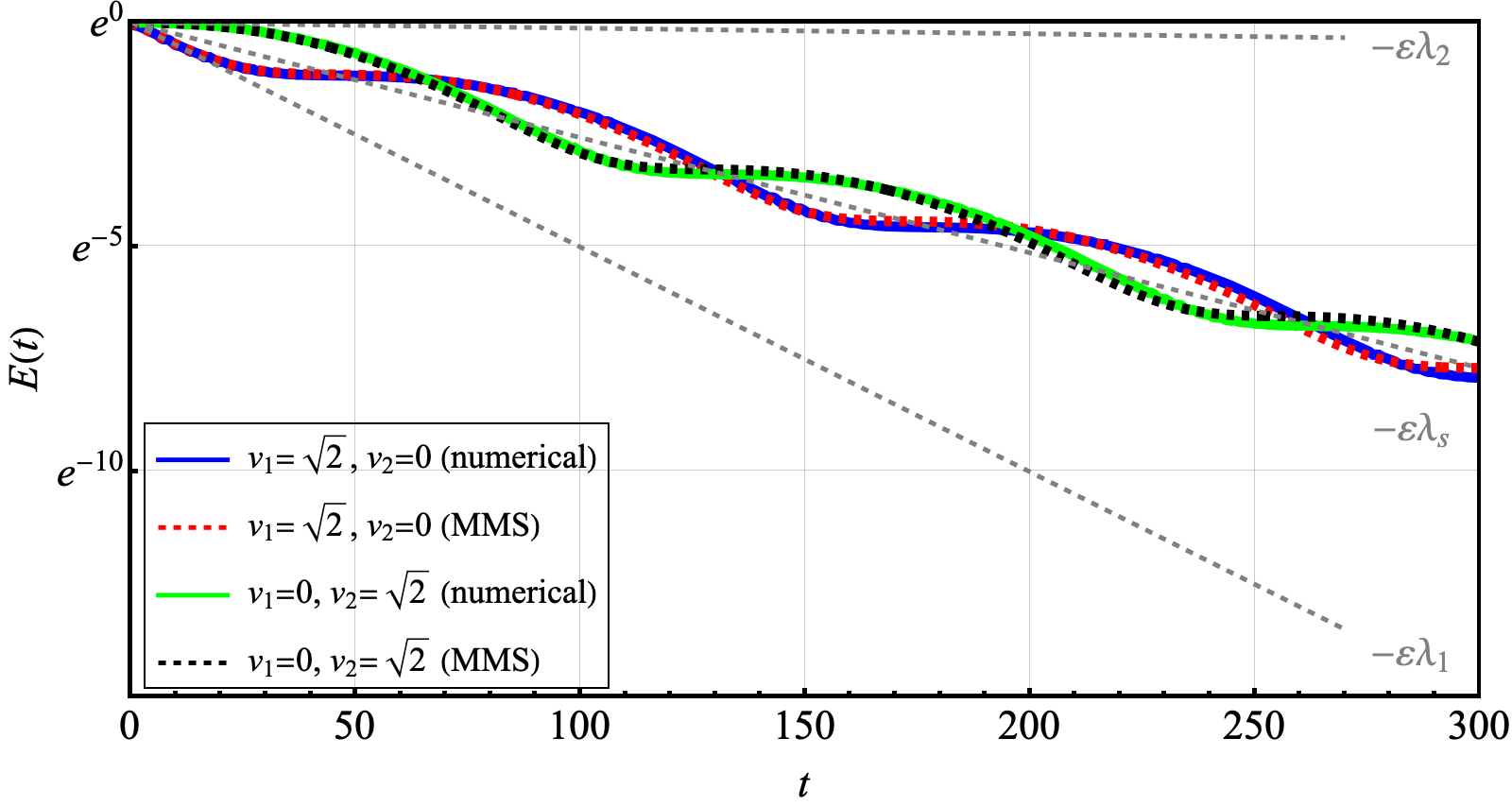}
        \caption{}
    \end{subfigure}
    \caption{Comparison of the logarithmic total energy decay $E(t)$ of the 2 DoF linear coupled oscillator~\eqref{eq:x1x2eom} for two impulse excitations, $v_1=\sqrt{2},v_2=0$ and $v_1=0,v_2=\sqrt{2}$, and for two values of the coupling stiffness: (a) $\beta=1.6<\beta_{\mathrm{crit}}$ (no modal interactions), and (b) $\beta=5.5>\beta_{\mathrm{crit}}$ (modal interactions), with $\varepsilon=0.01$, $\lambda_1=5$, and $\lambda_2=0.125$. The solid blue and green lines are obtained by numerical integration of Eq.~\eqref{eq:x1x2eom}, whereas the dashed red and black lines represent the leading-order analytical approximations of the total energy decay (cf. Eqs.~\eqref{eq:E1} and~\eqref{eq:E2}).}
    \label{fig:linenergies}
\end{figure}

Fig.~\ref{fig:linenergies} shows the total energy decay $E(t)$ (on a logarithmic scale) for two impulse excitations, $v_1=\sqrt{2},v_2=0$ and $v_1=0,v_2=\sqrt{2}$, and for two values of the coupling stiffness: (a) $\beta=1.6<\beta_{\mathrm{crit}}$ (no modal interactions), and (b) $\beta=5.5>\beta_{\mathrm{crit}}$ (modal interactions). Clearly, the action-angle formulation provides a good leading-order approximation (dashed lines) for the energy response, thereby confirming the analytical results presented above. For reference, the energy-decay slopes corresponding to the damping ratios $\varepsilon\lambda_1=0.05$ and $\varepsilon\lambda_2=0.00125$ ($\varepsilon=0.01$), and to the average ratio $\varepsilon\lambda_s=0.025625$ are shown as gray dashed lines. These damping ratios give, $\lambda_a=2.4375$, which is also the value of the critical coupling $\beta_{\mathrm{crit}}$ that separates the two dynamical regimes mentioned above. For coupling stiffness below the critical value, $\beta<\beta_{\mathrm{crit}}$ (left panel of Fig.~\ref{fig:linenergies}), the system exhibits two distinct decay rates and no modal interactions occur. Above the critical coupling, $\beta>\beta_{\mathrm{crit}}$ (right panel), the energy decays with the average decay rate, and intense modal interactions in the form of beat phenomena and energy exchanges between the two modes occur \cite{pinto2026understanding}. 

In synopsis, this section introduced a new asymptotic methodology based on action-angle transformations and the method of multiple scales to study modal interactions. This methodology will be applied in the next section to analyze modal interactions in the same system, but with a local strong stiffness nonlinearity of the third order.

\section{Analysis of modal interactions in the nonlinear system}
\label{sec:nonlin}
In this section, we consider the 2 DoF coupled oscillator when the left mass $m_1$ is grounded through a strongly nonlinear stiffness of the third degree, and set $\beta_1=0$. The corresponding equations of motion are now given by:
\begin{equation}
\begin{aligned}
 &\ddot{x}_1+\varepsilon\lambda_1 \dot{x}_1+\alpha x_1^3+\varepsilon\beta\left(x_1-x_2\right)=0 \\
&\ddot{x}_2+\varepsilon\lambda_2 \dot{x}_2+x_2+\varepsilon\beta\left(x_2-x_1\right)=0\\
       & x_1 (0)=x_2 (0)=0,\quad \dot{x}_1 (0)=v_1,\quad \dot{x}_2 (0)=v_2 \label{eq:nonlineom}
\end{aligned}
\end{equation}
As for the linear system~\eqref{eq:x1x2eom}, we investigate the free-decaying response of system~\eqref{eq:nonlineom} under impulsive excitation of either of the two masses, i.e., $v_1=\sqrt{2E_0},v_2=0$ and $v_1=0,v_2=\sqrt{2E_0}$, considering different values of the initial energy $E_0$, since in contrast to the linear system~\eqref{eq:x1x2eom} the dynamics of the nonlinear system~\eqref{eq:nonlineom} are energy-dependent. Moreover, we will consider the dynamics close to a 1:1 resonance manifold of the dynamics, i.e., only in the regime where the difference between the instantaneous frequencies of the two DoF is small, and 1:1 TRC occurs \cite{arnold1985mathematical,vakakis2008nonlinear}. Unlike the linear system~\eqref{eq:x1x2eom}, the stiffness symmetry in system~\eqref{eq:nonlineom} is broken due to the nonlinear stiffness of the left DoF and also that now $\beta_1=0$. Consequently, the responses of the two DoFs depend not only on the location of the initial excitation but also on the relation between the non-proportional damping coefficients, i.e., $\lambda_1>\lambda_2$, $\lambda_1=\lambda_2$ or $\lambda_2>\lambda_1$; each of these cases results in quantitatively different energy-transfers between the two DoFs, and thus, energy decay characteristics, even if the underlying 1:1 TRC remains qualitatively unchanged.

We introduce the action-angle formulation for the two uncoupled, undamped and unforced left (nonlinear) and right (linear) oscillators \cite{vakakis2001energy},
\begin{equation}
    \begin{aligned}
        x_1(t)&=\Lambda  I_1^{1/3}(t) \mathrm{cn}\left(\frac{2 K(m) \theta_1(t)}{\pi}, m\right) \\
\dot{x}_1 (t)&=-\Lambda I_1^{1/3}(t)   \omega_1\left(I_1(t)\right) \frac{2 K(m)}{\pi}
\mathrm{sn}\left(\frac{2 K(m) \theta_1(t)}{\pi}, m\right)\mathrm{dn}\left(\frac{2 K(m) \theta_1(t)}{\pi}, m\right)\\
        x_2(t)&=\sqrt{2I_2(t)}\sin\theta_2(t)\\
        \dot{x}_2(t)&=\sqrt{2I_2(t)}\cos\theta_2(t) \label{eq:actionangles}
    \end{aligned}
\end{equation}
where $\mathrm{sn}\left(\frac{2 K(m) \theta_1(t)}{\pi}, m\right)$ and $\mathrm{dn}\left(\frac{2 K(m) \theta_1(t)}{\pi}, m\right)$ are the Jacobi elliptic sine and delta functions, respectively, and $K$ is the complete elliptic integral of the first kind with an elliptic parameter $m = 1/2$. In addition, $\Lambda=\left(\frac{1}{4\alpha}\right)^{1/6}\left(\frac{3\pi}{K\left(m\right)}\right)^{1/3}$ and $\omega_1(I_1)=\left(\frac{3\pi^4\alpha}{16K^4\left(m\right)}\right)^{1/3} I_1^{1/3}$. For detailed derivation see the supplementary material in the \hyperref[appendix]{Appendix}. 

Writing the two second-order nonlinear differential equations of motion~\eqref{eq:nonlineom} as a set of four first order nonlinear differential equations, and introducing the action-angle transformations~\eqref{eq:actionangles}, we express the governing equations as,
\begin{subequations}\label{eq:angles}
\begin{align}
\dot{I}_1(t)&=-\varepsilon
\notag\Bigg(\lambda_1\Lambda^2 I_1^{2/3}(t)  \omega_1(I_1(t))\frac{4K^2(m)}{\pi^2}
\mathrm{sn}^2\left(\frac{2 K(m) \theta_1(t)}{\pi}, m\right)\mathrm{dn}^2\left(\frac{2 K(m) \theta_1(t)}{\pi}, m\right)- \\ \notag& -\beta\Lambda^2 I_1^{2/3}(t)\frac{2K(m)}{\pi}\mathrm{cn}\left(\frac{2 K(m) \theta_1(t)}{\pi}, m\right)\mathrm{sn}\left(\frac{2 K(m) \theta_1(t)}{\pi}, m\right)\times \\ \notag& \times\mathrm{dn}\left(\frac{2 K(m) \theta_1(t)}{\pi}, m\right)+\beta\Lambda I_1^{1/3}(t)\frac{2K(m)}{\pi}\sqrt{2I_2(t)}\times \\ &\times\mathrm{sn}\left(\frac{2 K(m) \theta_1(t)}{\pi}, m\right)\mathrm{dn}\left(\frac{2 K(m) \theta_1(t)}{\pi}, m\right)\sin\theta_2(t)\Bigg) \\
     \notag   \dot{I}_2(t)&=-\varepsilon\Bigg(2\lambda_2I_2(t)\cos^2\theta_2(t)+2\beta I_2(t)\sin\theta_2(t)\cos\theta_2(t)-\\ \notag&-\beta \Lambda I_1^{1/3}(t)\sqrt{2I_2(t)}\mathrm{cn}\left(\frac{2 K(m) \theta_1(t)}{\pi}, m\right)\cos\theta_2(t)\Bigg)
\end{align}


\begin{align}
\notag\dot{\theta}_1(t)&=\omega(I_1(t))-\varepsilon
\Bigg(\frac{\lambda_1\Lambda^2}{3} \frac{ \omega_1(I_1(t))}{I_1^{1/3}(t) }\frac{2K(1/2)}{\pi}
\mathrm{cn}\left(\frac{2 K(m) \theta_1(t)}{\pi}, m\right)\mathrm{sn}\left(\frac{2 K(m) \theta_1(t)}{\pi}, m\right)\times \\ &\qquad \qquad \times\mathrm{dn}\left(\frac{2 K(m) \theta_1(t)}{\pi}, m\right) -\frac{\beta\Lambda^2}{3I_1^{1/3}(t)} \mathrm{cn}^2\left(\frac{2 K(m) \theta_1(t)}{\pi}, m\right)+ \\ \notag &\qquad \qquad +\beta\Lambda\frac{\sqrt{2I_2(t)}}{3I_1^{2/3}(t)}  \mathrm{cn}\left(\frac{2 K(m) \theta_1(t)}{\pi}, m\right)\sin\theta_2(t)\Bigg)\\
\notag \dot{\theta}_2(t)&=1+\varepsilon\Bigg(\lambda_2\cos\theta_2(t)\sin\theta_2(t)+\beta \sin^2\theta_2(t)-\beta \Lambda\frac{I_1^{1/3}(t)}{\sqrt{2I_2(t)}}\mathrm{cn}\left(\frac{2 K(m) \theta_1(t)}{\pi}, m\right)\sin\theta_2(t)\Bigg)
\end{align}
\end{subequations}
with initial conditions:
\begin{equation}
    \begin{aligned}
    I_1(0)=\left(\frac{2v_1^2}{\alpha\Lambda^4}\right)^{3/4}&,\quad I_2(0)=\frac{v_2^2}{2}\\
    \theta_1(0)=\left\{\begin{array}[c]{c}
        -\frac{\pi}{2},\quad v_1\neq0, v_2=0\\
        -\pi,\quad v_1=0, v_2\neq0\\
    \end{array}\right.&,\quad \theta_2(0)=\left\{\begin{array}[c]{c}
        -\frac{\pi}{2},\quad v_1\neq0, v_2=0\\
        0,\quad v_1=0, v_2\neq0\\
    \end{array}\right.
\end{aligned} \label{eq:nonlinactini}
\end{equation}

It follows from Eq.~\eqref{eq:angles} that the nonlinear system of coupled oscillators~\eqref{eq:nonlineom} has two fast frequencies, $\dot{\theta}_1(t)\approx\omega_1(I_1(t))$ and $\dot{\theta}_2(t)\approx1$. In contrast to the linear oscillator~\eqref{eq:x1x2eom}, the frequency of the nonlinear DoF depends on its respective action, and for small values of the parameter $\varepsilon$ varies slowly in time. Consequently,
in the presence of weak coupling, $\varepsilon\beta$, and damping, $\varepsilon\lambda_1,\varepsilon\lambda_2$, a 1:1 TRC \cite{arnold1985mathematical,vakakis2008nonlinear}, analogous to that observed in the linearly coupled oscillator~\eqref{eq:x1x2eom}, can occur when the action $I_1(t)$ reaches a critical level, $I_{1,\mathrm{cr}}$, so that the following 1:1 resonance condition is approximately satisfied,
\begin{equation}
    \omega_1(I_{1,\mathrm{cr}})\approx1\label{eq:cond}
\end{equation}
from which the critical value of the action is computed as:
\begin{equation}
    I_{1,\mathrm{cr}}=\frac{16K^4\left(m\right)}{3\pi^4\alpha} \label{eq:actcrit}
\end{equation}
Note that the critical value for the action, $I_{1,\mathrm{cr}}$, is inversely proportional to the nonlinear stiffness $\alpha$. We emphasize that Eq.~\eqref{eq:cond} is a \emph{necessary but not sufficient condition} for the realization of 1:1 TRC in the dynamics. If 1:1 TRC is indeed realized, we anticipate that modal interactions similar to the linear case studied in Section~\ref{sec:lin} would hold, resulting in strongly nonlinear energy transfers between the nonlinear normal modes \cite{vakakis1996normal} of system~\eqref{eq:nonlineom}; however, such nonlinear modal interactions should occur only over the finite time duration of capture of the dynamics in the neighborhood of the 1:1 resonance manifold -- during which condition~\eqref{eq:cond} holds, and cease to exist after the transient dynamics escape from 1:1 TRC. In the following analysis we study the necessary and sufficient conditions for this transient resonance to occur, and analyze asymptotically (in the limit of small $\varepsilon$) the transient dynamics in that regime.

\subsection{Necessary and sufficient conditions for 1:1 transient resonance capture (TRC)}

\begin{figure}[ptb]
    \centering
    \includegraphics[width=.9\linewidth]{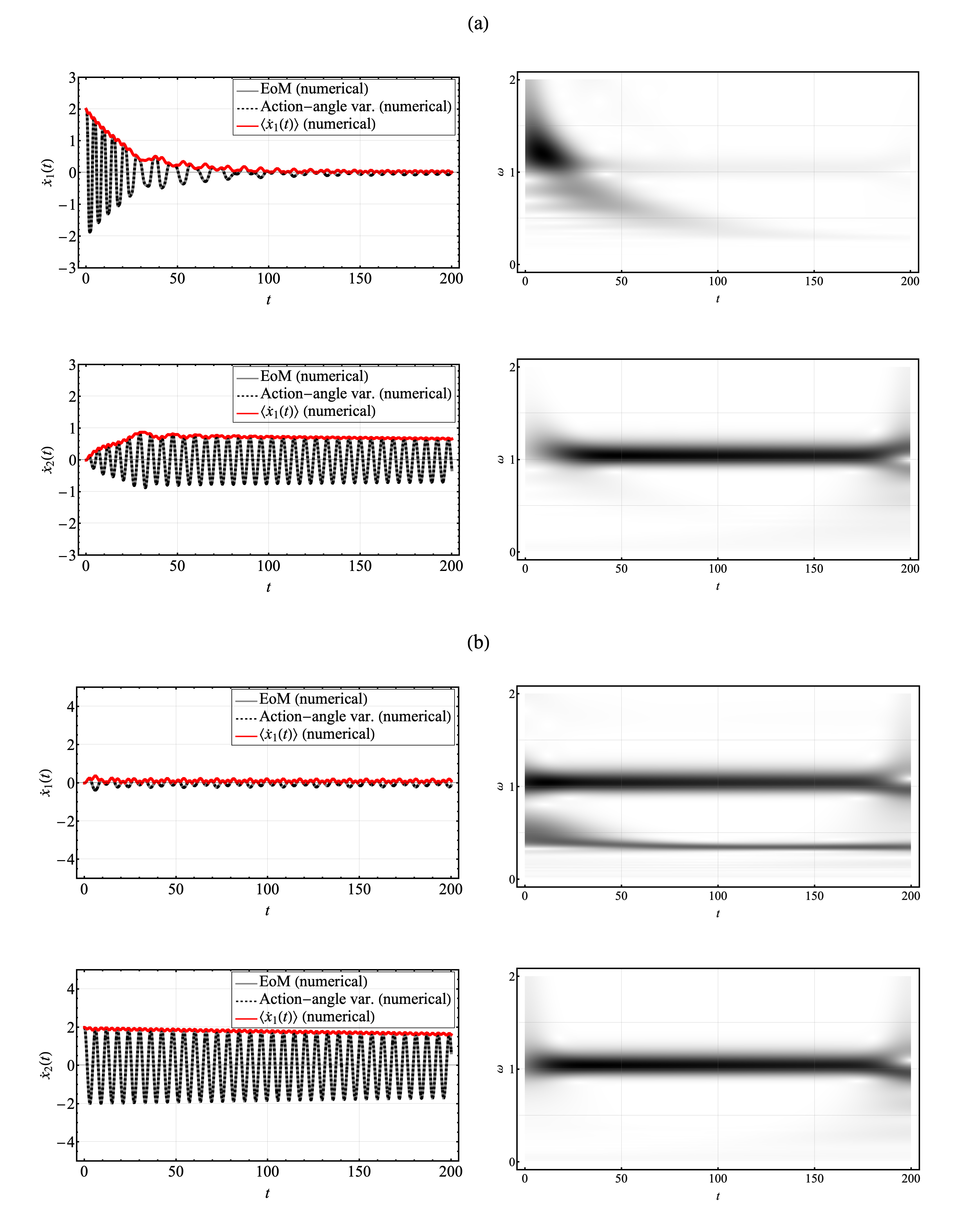}
    \caption{Comparison of the velocity time series and corresponding wavelet transforms of the two DoFs, $m_1$ and $m_2$, with parameters $\varepsilon=0.01$, $\lambda_1=5$, $\lambda_2=0.125$, $\alpha=1$, and a coupling stiffness $\beta=8$. The system is subjected to an excitation with impulse intensities (a) $v_1=\sqrt{2E_0},v_2=0$ (no 1:1 TRC) and (b) $v_1=0,v_2=\sqrt{2E_0}$ (no 1:1 TRC), for initial energy $E_0=2.0$. The solid gray lines are the numerical solutions obtained by integrating the equations of motion~\eqref{eq:nonlineom}, while the dashed, black lines denote the velocities computed by~\eqref{eq:actionangles}, obtained through numerical integration of Eq.~\eqref{eq:angles}. The corresponding envelopes governed by Eq.~\eqref{eq:nonlinenv} are shown as solid red curves.}
    \label{fig:nonlinvelmass2}
\end{figure}

The first issue in the analysis is to determine the parameter values or initial energies for which the necessary condition \eqref{eq:cond} is satisfied for the two impulse excitation cases. For the first excitation case, $v_1=\sqrt{2E_0},v_2=0$, using the initial condition~\eqref{eq:nonlinactini} for the action $I_1(t)$ and the critical action $I_{1,\mathrm{cr}}$ from Eq.~\eqref{eq:actcrit} the minimal necessary initial energy $E_0$ is,
\begin{equation}
   I_1(0)\geq I_{1,\mathrm{cr}} \quad\rightarrow\quad   E_0\geq \frac{\alpha\Lambda^4}{4}I_{1,\mathrm{cr}}^{{4/3}}= \frac{4K^4\left(m\right)}{ \pi^4\alpha} \label{eq:cond1}
\end{equation}
and only depends on the nonlinear stiffness $\alpha$. In the following, we apply an excitation $v_1=\sqrt{2E_0},v_2=0$ with initial energy $E_0=2.0$ (well above the threshold
given by Eq. \eqref{eq:cond1}) to system~\eqref{eq:nonlineom} with a coupling stiffness of $\varepsilon\beta=0.08$ ($\varepsilon=0.01$), for damping coefficients $\lambda_1=5,\lambda_2=0.125$. Fig.~\ref{fig:nonlinvelmass2} (a) shows the velocity time series and corresponding wavelet transforms of the two DoFs, $m_1$ and $m_2$. The solid gray lines depict the exact solutions obtained by numerically integrating the equations of motion~\eqref{eq:nonlineom}, while the dashed, black lines denote the velocities given by~\eqref{eq:actionangles}, obtained through numerical integration of  Eq.~\eqref{eq:angles} as well. The corresponding velocity envelopes, from Eq.~\eqref{eq:actionangles} are:
\begin{equation}
    \begin{aligned}
 \langle\dot{x}_1 (t)\rangle&=-\frac{1}{\sqrt{2}}\Lambda I_1^{1/3}(t)   \omega_1\left(I_1(t)\right) \frac{2 K(m)}{\pi}
\\
      \langle    \dot{x}_2(t)\rangle&=\sqrt{2I_2(t)} \label{eq:nonlinenv}
    \end{aligned}
\end{equation}
It follows from Fig.~\ref{fig:nonlinvelmass2} (a) that, even though condition~\eqref{eq:cond1} is satisfied for $E_0=2.0$, no 1:1 TRC occurs, highlighting the fact that~\eqref{eq:cond1} represents the necessary but not sufficient condition for the realization of 1:1 TRC. When the nonlinear DoF is heavily damped and the initial impulse is applied to this DoF, its action decays past $I_{1,\mathrm{cr}}$ before the weak coupling can transfer enough energy to the linear DoF for a resonant interaction to occur between the two DoFs. This implies the necessity for a sufficient condition for TRC, which we address later.

In comparison to the first excitation case, the second excitation case $v_1=0,v_2=\sqrt{2E_0}$, shown in Fig.~\ref{fig:nonlinvelmass2} (b), is more involved. In this case, the initial energy $E_0=2.0$ is induced to the linear DoF, therefore the initial action of the nonlinear DoF is zero. This means, that the energy is transferred from the linear DoF to the nonlinear DoF through the weak coupling, $\varepsilon\beta$, such that the action $I_1(t)$ reaches the critical value $I_{1,\mathrm{cr}}$, only when $E_0$ is sufficiently large to drive the nonlinear DoF into 1:1 resonance capture. For parameter values $\varepsilon=0.01$, $\alpha=1$, $\beta=8$, $\lambda_1=5$, and $\lambda_2=0.125$, the initial energy $E_0=2.0$ proves to be insufficient as the resulting time series and their wavelet transforms show in Fig.~\ref{fig:nonlinvelmass2} (b). From Fig.~\ref{fig:nonlinvelmass2} (b) follows that no TRC can occur. The wavelet transform of the response of the nonlinear DoF shows that its oscillation frequency remains below unity. The strong frequency component appearing around unity is attributed to the excitation transmitted through the coupling. Based on these, an investigation of the early time interaction between the two DoFs is necessary. Eq.~\eqref{eq:angles} represents a two-frequency dynamical system for which a two-frequency averaging is applicable. However, as shown in \cite{vakakis2001energy} such a straightforward averaging procedure fails when the conditions of 1:1 TRC exist and energy exchanges between the modes take place. Hence, we determine later numerically the minimum initial energy $E_0$ required in the second excitation case for the action $I_1(t)$ to reach the critical value $I_{1,\mathrm{cr}}$, considering different values of nonlinear stiffness $\alpha$, coupling stiffness $\beta$, and damping ratio $\lambda_1$.

\begin{figure}[ptb]
    \centering
    \includegraphics[width=0.9\linewidth]{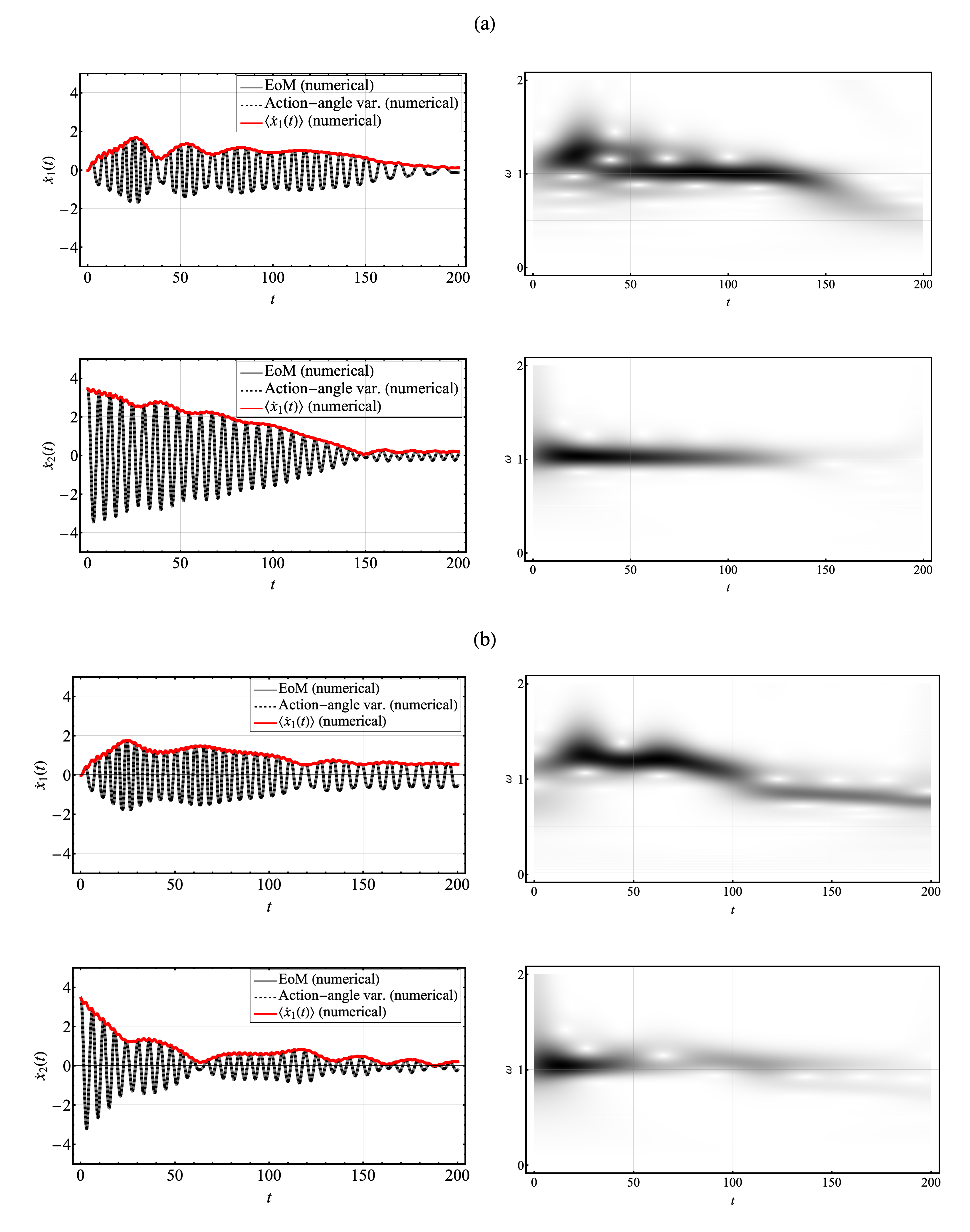}
    \caption{Comparison of the velocity time series and corresponding wavelet transforms of the two DoFs, $m_1$ and $m_2$, with parameters $\varepsilon=0.01$, $\alpha=1$, and a coupling stiffness $\beta=8$. The damping ratios are (a) $\lambda_1=5$, $\lambda_2=0.125$ and (b) $\lambda_1=0.125$, $\lambda_2=5$. The system is subjected to an excitation with impulse intensities $v_1=0,v_2=\sqrt{2E_0}$, for initial energy $E_0=6.0$ (1:1 TRC). The solid gray lines are the numerical solutions obtained by integrating the equations of motion~\eqref{eq:nonlineom}, while the dashed, black lines denote the velocities computed by~\eqref{eq:actionangles}, obtained through numerical integration of Eq.~\eqref{eq:angles}. The corresponding envelopes governed by Eq.~\eqref{eq:nonlinenv} are shown as solid red curves.}
    \label{fig:nonlinvelmass2_2}
\end{figure}

Fig.~\ref{fig:nonlinvelmass2_2} (a) show the case when the initial energy is increased to a level ($E_0=6.0$) within the region where 1:1 TRC can occur (the necessary condition is met) the linear DoF can sufficiently excite the nonlinear DoF through the coupling, allowing its frequency to reach unity, and leading to recurring beat phenomena. The corresponding oscillatory responses possess a fast frequency equal to unity and a slow frequency associated with the modulation of the oscillation envelopes. These envelopes reflect the modal interactions between the two nonlinear modes of the system, resulting in more effective energy dissipation.

Fig.~\ref{fig:nonlinvelmass2_2} (b) demonstrates a resonance capture in system~\eqref{eq:nonlineom} with identical parameters to the previous simulation except for the damping coefficients, $\lambda_1=0.125$ and $\lambda_2=5$, i.e., the values of $\lambda_1$ and $\lambda_2$ were exchanged. The resulting dynamics resembles the case $\lambda_1>\lambda_2$, confirming that the relative magnitudes of the damping coefficients (damping distribution) do not affect whether 1:1 TRC occurs.

The second issue, namely, the condition for the existence of an attracting resonant manifold in the phase space is now determined. As mentioned above, even if the initial impulsive energy $E_0$, is sufficiently large for the action $I_1(t)$ to reach $I_{1,\mathrm{cr}}$, the existence of a resonant manifold in the phase space must be investigated separately. Accordingly, we restrict our analysis to a parameter regime in which 1:1 TRC can occur in system~\eqref{eq:nonlineom}. To analyze the dynamics in the vicinity of the resonant manifold, we introduce the phase difference,
\begin{equation}
    \psi(t)=\theta_1(t)-\theta_2(t)\label{eq:nonlinpsi}
\end{equation}
which serves as the slow angle associated with 1:1 TRC. Using Eq.~\eqref{eq:angles}, the evolution of the slow angle~\eqref{eq:nonlinpsi} is,
\begin{equation}
    \begin{aligned}
    \dot{\psi}(t)&=\omega(I_1(t))-1-\varepsilon\Bigg(\frac{\lambda_1\Lambda^2 }{3I_1^{1/3}(t)}  \omega_1(I_1(t))\frac{2K(m)}{\pi}
\mathrm{cn}\left(\frac{2 K(m) \phi(t)}{\pi}, m\right)\times \\ &\times\mathrm{sn}\left(\frac{2 K(m) \phi(t)}{\pi}, m\right) \mathrm{dn}\left(\frac{2 K(m) \phi(t)}{\pi}, m\right)- \frac{\beta\Lambda^2 }{3I_1^{1/3}(t)}\mathrm{cn}^2\left(\frac{2 K(m) \phi(t)}{\pi}, m\right)+ \\ &+\beta\Lambda\frac{\sqrt{2I_2(t)}}{3I_1^{2/3}(t) } \mathrm{cn}\left(\frac{2 K(m) \phi(t)}{\pi}, m\right)\sin\theta_2(t)+\lambda_2\cos\theta_2(t)\sin\theta_2(t)+\\ &+\beta \sin^2\theta_2(t)-\beta \Lambda\frac{ I_1^{1/3}(t)}{\sqrt{2I_2(t)}}\mathrm{cn}\left(\frac{2 K(m) \phi(t)}{\pi}, m\right)\sin\theta_2(t)\Bigg) \label{eq:slowDiff}
    \end{aligned}
\end{equation}
where,
\begin{equation}
    \phi(t)=\psi(t)+\theta_2(t)
\end{equation}

As discussed above, the necessary condition for 1:1 resonance occurring in system~\eqref{eq:nonlineom} is:
\begin{equation}
\omega_1(I_{1,\mathrm{cr}})\approx1 \label{eq:condnew}
\end{equation}
Under this condition, $\psi(t)$ is a slow angle with $\theta_2(t)$ being the fast angle, which has time-like behavior (cf. Eq.~\eqref{eq:angles}). Accordingly, we can replace the time variable with the phase angle,
\begin{equation}
    t=\theta_2(t)+\mathcal{O}(\varepsilon) \label{eq:dep}
\end{equation}
and study the local dynamics of the system in a neighborhood of the resonant manifold, by introducing the new (boundary-layer) local coordinates \cite{vakakis2001energy},
\begin{equation}
    I_1(t)=I_{1,\mathrm{cr}}+\sqrt{\varepsilon}\xi(t),\quad I_2(t)=\eta(t)
\end{equation}
in terms of which we obtain the following equations governing the local dynamics
\begin{equation}
    \begin{aligned}
        \dot{\xi}(t)&=-\sqrt{\varepsilon}\Bigg(\lambda_1\Lambda^2 I_{1,\mathrm{cr}}^{2/3}  \omega_1(I_{1,\mathrm{cr}})\frac{4K^2(1/2)}{\pi^2}
\mathrm{sn}^2\left(\frac{2K(m)}{\pi}\phi(t),m\right)\mathrm{dn}^2\left(\frac{2K(m)}{\pi}\phi(t),m\right)- \\ & -\beta\Lambda^2 I_{1,\mathrm{cr}}^{2/3}\frac{2K(m)}{\pi}\mathrm{cn}\left(\frac{2K(m)}{\pi}\phi(t),m\right)\mathrm{sn}\left(\frac{2K(m)}{\pi}\phi(t),m\right)\mathrm{dn}\left(\frac{2K(m)}{\pi}\phi(t),m\right)+\\ &+\beta\Lambda I_{1,\mathrm{cr}}^{1/3}\frac{2K(m)}{\pi}\sqrt{2\eta(t)}\mathrm{sn}\left(\frac{2K(m)}{\pi}\phi(t),m\right)\mathrm{dn}\left(\frac{2K(m)}{\pi}\phi(t),m\right)\sin\theta_2(t)\Bigg)+\mathcal{O}(\varepsilon) \\
        \dot{\eta}(t)&=-\varepsilon\Bigg(2\lambda_2\eta(t)\cos^2\theta_2(t)+2\beta \eta(t)\sin\theta_2(t)\cos\theta_2(t)-\\ &\qquad\qquad-\beta \Lambda I_{1,\mathrm{cr}}^{{1/3}}\sqrt{2\eta(t)}\mathrm{cn}\left(\frac{2K(m)}{\pi}\phi(t),m\right)\cos\theta_2(t)\Bigg)+\mathcal{O}(\varepsilon^{3/2}) \\
    \dot{\psi}(t)&=\sqrt{\varepsilon}\frac{\xi(t)}{3I_{1,\mathrm{cr}}^{{2/3}}}\left(\frac{3\pi^4\alpha}{16K^4(m)}\right)^{1/3}+\mathcal{O}(\varepsilon)\label{eq:inv}
    \end{aligned}
\end{equation}
 Applying the method of multiple scales through the introduction of the fast and slow scales,  
\begin{equation}
    T_0=\theta_2,\qquad T_1=\sqrt{\varepsilon} \theta_2
\end{equation}
and considering the asymptotic solutions of Eq.~\eqref{eq:inv} in the form,
\begin{equation}
    \begin{aligned}
        \xi(\theta_2)=\xi_0(T_0,T_1)+\sqrt{\varepsilon} \xi_1(T_0,T_1)+\mathcal{O}(\varepsilon^{3/2})\\
        \eta(\theta_2)=\eta_0(T_0,T_1)+\sqrt{\varepsilon} \eta_1(T_0,T_1)+\mathcal{O}(\varepsilon^{3/2})\\
        \psi(\theta_2)=\psi_0(T_0,T_1)+\sqrt{\varepsilon} \psi_1(T_0,T_1)+\mathcal{O}(\varepsilon^{3/2})\label{eq:appr}
    \end{aligned}
\end{equation}
we obtain a series of subproblems at different orders of the small parameter $\varepsilon$. The leading-order approximations at order $\varepsilon^0$ are given by, 
\begin{equation}
    \begin{aligned}
       \frac{\partial \xi_0}{\partial T_0} &=0 \qquad \rightarrow \qquad  \xi_0(T_0,T_1)=A_0(T_1)\\
   \frac{\partial \eta_0}{\partial T_0} &=0\qquad  \rightarrow \qquad \eta_0(T_0,T_1)=B_0(T_1)\\
   \frac{\partial \psi_0}{\partial T_0} &=0\qquad  \rightarrow \qquad \psi_0(T_0,T_1)=C_0(T_1) \label{eq:zeroth}
    \end{aligned}
\end{equation}
Moreover, the $\varepsilon^{1/2}$ approximations are governed by the following set of equations:
\begin{equation}
    \begin{aligned}
       \frac{\partial \xi_1}{\partial T_0} &=-A_0^\prime(T_1)-\\ & -\Bigg(\lambda_1\Lambda^2 I_{1,\mathrm{cr}}^{2/3}  \frac{4K^2(m)}{\pi^2}
\mathrm{sn}^2\left(\frac{2K(m)}{\pi}(C_0(T_1)+T_0),m\right)\mathrm{dn}^2\left(\frac{2K(m)}{\pi}(C_0(T_1)+T_0),m\right)- \\ & -\beta\Lambda^2 I_{1,\mathrm{cr}}^{2/3}\frac{2K(m)}{\pi}\mathrm{cn}\left(\frac{2K(m)}{\pi}(C_0(T_1)+T_0),m\right)\mathrm{sn}\left(\frac{2K(m)}{\pi}(C_0(T_1)+T_0),m\right)\\ &\mathrm{dn}\left(\frac{2K(m)}{\pi}(C_0(T_1)+T_0),m\right)+\beta\Lambda I_{1,\mathrm{cr}}^{1/3}\frac{2K(1/2)}{\pi}\sqrt{2\eta_0}\mathrm{sn}\left(\frac{2K(m)}{\pi}(C_0(T_1)+T_0),m\right)\\ &\mathrm{dn}\left(\frac{2K(m)}{\pi}(C_0(T_1)+T_0),m\right)\sin T_0\Bigg)\\
   \frac{\partial \eta_1}{\partial T_0} &=-B_0^\prime(T_1)\\
   \frac{\partial \psi_1}{\partial T_0} &=-C_0^\prime(T_1) +\frac{A_0(T_1)}{3I_{1,\mathrm{cr}}^{{2/3}}}\left(\frac{3\pi^4\alpha}{16K^4(m)}\right)^{1/3}\label{eq:firstapprox}
    \end{aligned}
\end{equation}
To eliminate the secular terms we impose the solvability conditions. Namely, the average over one $2\pi$-period in $T_0$ of the right-hand sides of Eq.~\eqref{eq:firstapprox} have to be set equal to zero, giving the following slow-flow equations at this order of approximation:
\begin{equation}
    \begin{aligned}
       A_0^\prime(T_1)&=-\frac{\lambda_1\Lambda^2 I_{1,\mathrm{cr}}^{2/3}}{3}  \frac{4K^2(m)}{\pi^2} -\beta\Lambda I_{1,\mathrm{cr}}^{1/3}\frac{2K(m)}{\pi}\sqrt{2B_0}\frac{\sqrt{2}\pi^2e^{-\pi/2}}{2K^2(m)(1+e^{-\pi})}\cos C_0(T_1)\\
       B_0^\prime(T_1)&=0\\
   C_0^\prime(T_1)&=\frac{A_0(T_1)}{3I_{1,\mathrm{cr}}^{{2/3}}}\left(\frac{3\pi^4\alpha}{16K^4(m)}\right)^{1/3} \label{eq:pend}
    \end{aligned}
\end{equation}
These slow-flow equations govern the evolution of the slow actions and slow angle of the dynamics in the regime of 1:1 TRC. Note that the action $B_0$ of the linear DoF is conserved at this order, while action $A_0$ corresponding to the nonlinear DoF and the slow angle $C_0$ form a separate two-dimensional subsystem governed by,
\begin{equation}
    C_0^{\prime\prime}(T_1)+\frac{\pi^4 e^{-\pi/2}}{4K^4(m)(1+e^{-\pi})}\beta\sqrt{\alpha B_0}\cos C_0(T_1)=-\frac{1}{3}\lambda_1 \label{eq:pend2}
\end{equation}
or, numerically evaluating the parameter-independent coefficient:
\begin{equation}
    C_0^{\prime\prime}(T_1)+0.410647 \beta\sqrt{\alpha B_0}\cos C_0(T_1)=-\frac{1}{3}\lambda_1 \label{eq:numerical}
\end{equation}
Thus, the slow angle dynamics reduce to a pendulum equation subject to a constant "torque" (the term on the right-hand side). The phase-locked state, when 1:1 TRC occurs, is obtained from the fixed-point conditions,
\begin{equation}
    C_{0,\mathrm{e}}^\prime\equiv0,\qquad \cos C_{0,\mathrm{e}}\equiv-\frac{0.811727\lambda_1}{\beta\sqrt{\alpha B_0}}
\end{equation}
which exist iff:
\begin{equation}
    \Bigg\vert \frac{0.811727\lambda_1}{\beta\sqrt{\alpha B_0}}\Bigg\vert\leq1
\end{equation}
Thus, a condition for the existence of resonant manifold in terms of the action (energy) of the linear DoF is,
\begin{equation}
  B_0\geq  \left( \frac{0.811727\lambda_1}{\beta\sqrt{\alpha}}\right)^2 \label{eq:cond2}
\end{equation}
This complements the previous necessary condition Eq.~\eqref{eq:condnew} and together they provide the necessary and sufficient conditions for 1:1 TRC.

\begin{figure}[ptb]
    \centering
    \begin{subfigure}{0.45\textwidth}
        \centering
        \includegraphics[width=\linewidth]{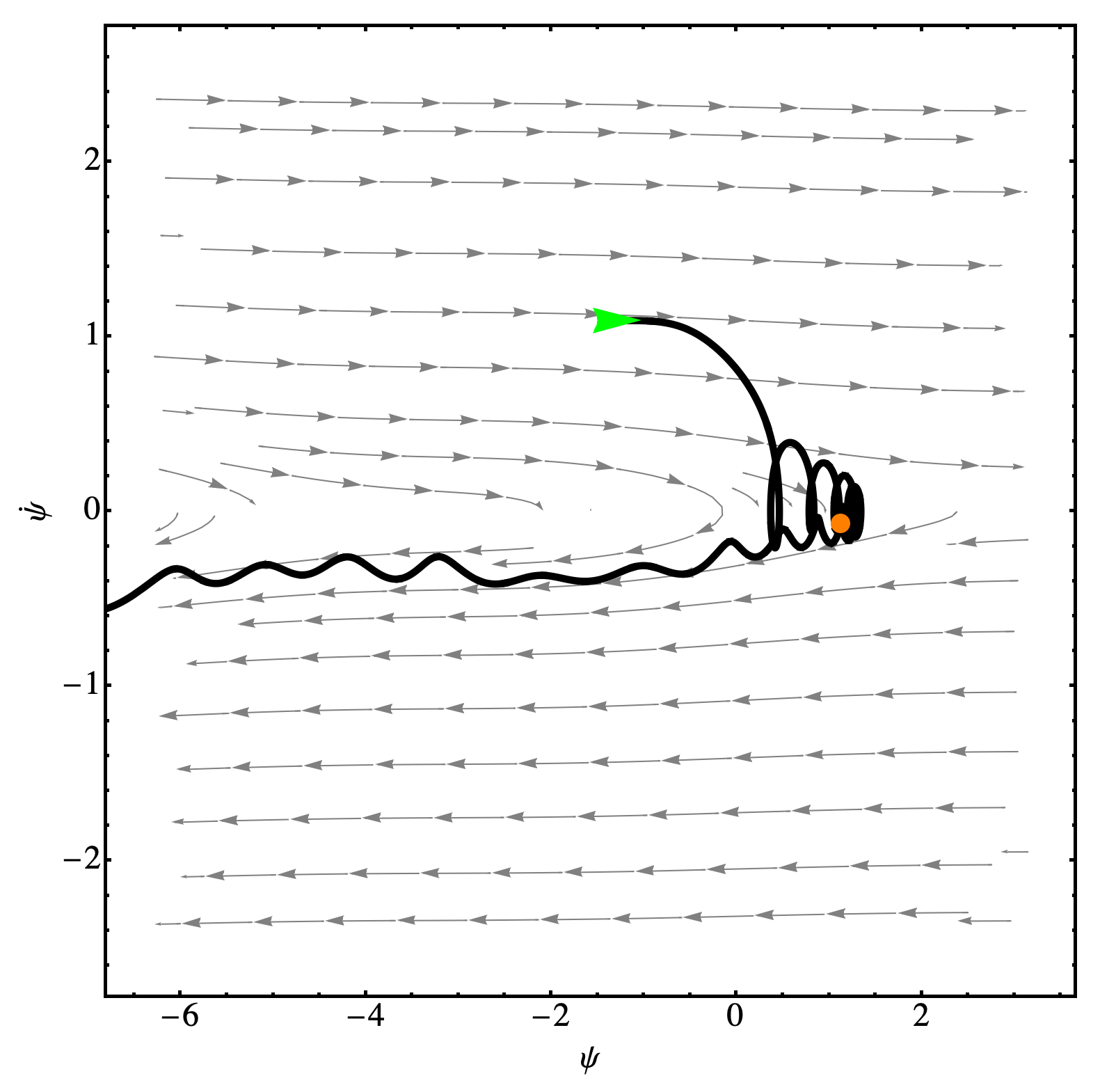}
        \caption{}
    \end{subfigure}
    \hfill
    \begin{subfigure}{0.45\textwidth}
        \centering
        \includegraphics[width=\linewidth]{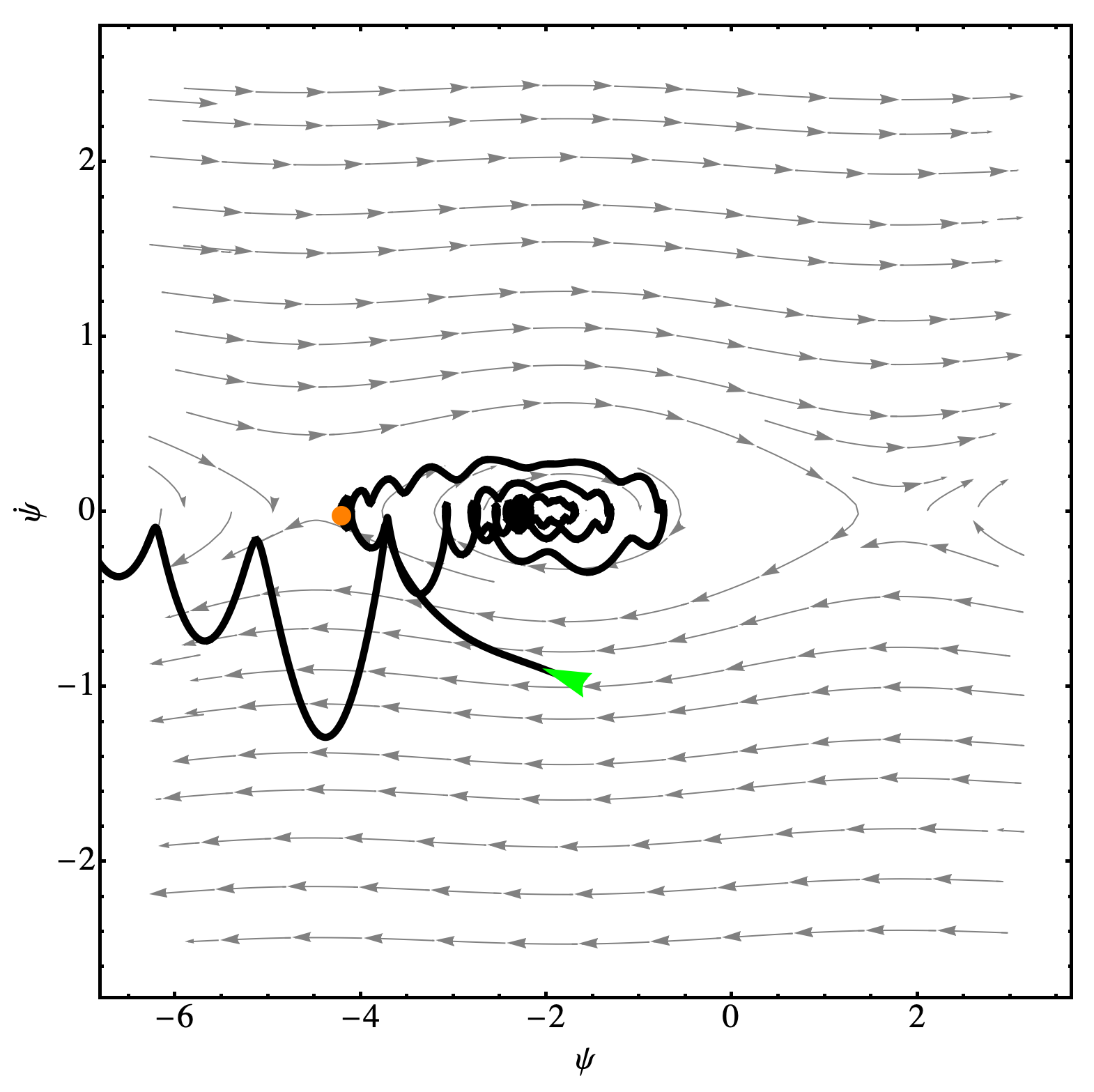}
        \caption{}
    \end{subfigure}

    \vspace{0.25cm}

    \begin{subfigure}{0.45\textwidth}
        \centering
        \includegraphics[width=\linewidth]{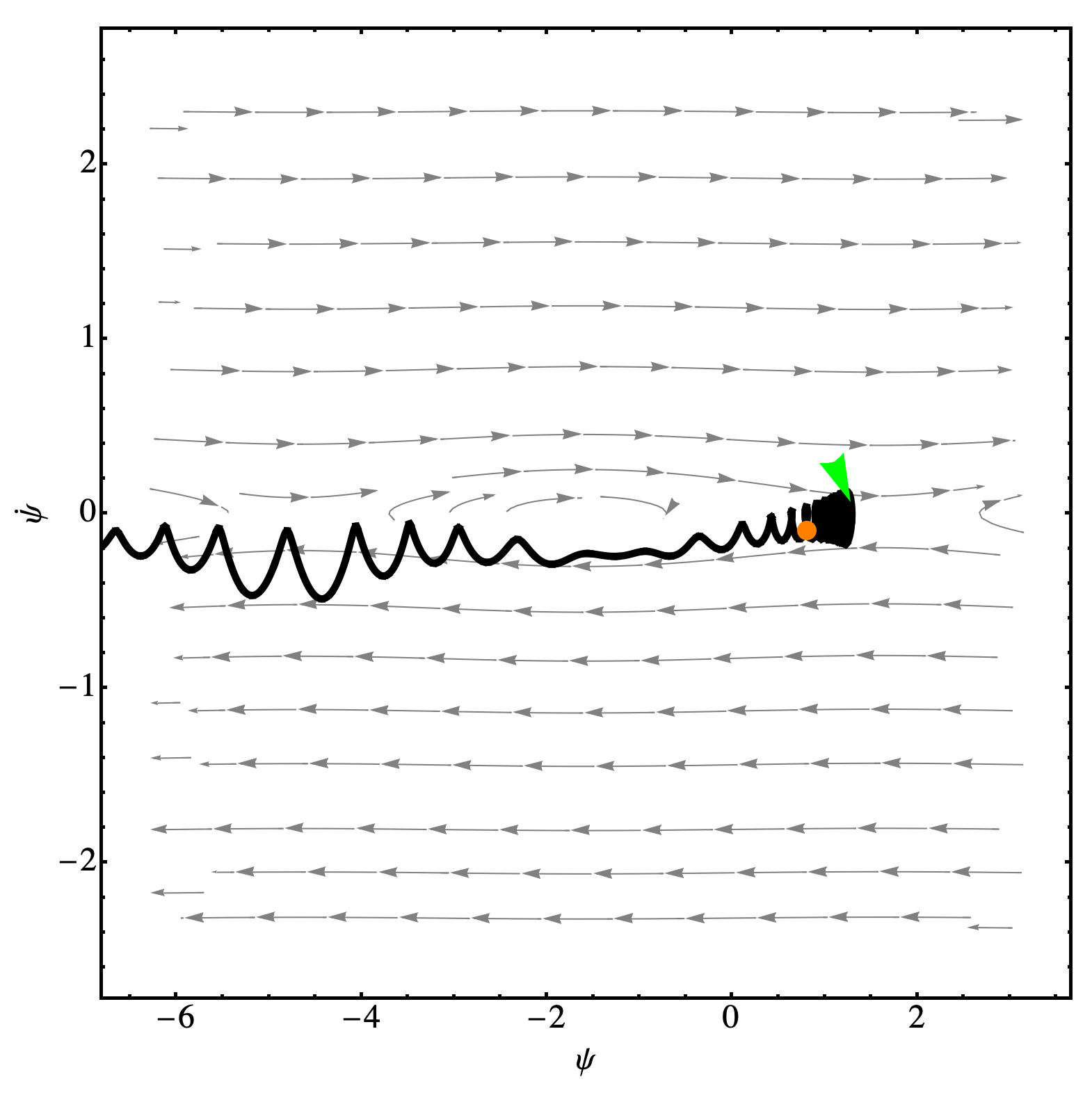}
        \caption{}
    \end{subfigure}
    \hfill
    \begin{subfigure}{0.45\textwidth}
        \centering
        \includegraphics[width=\linewidth]{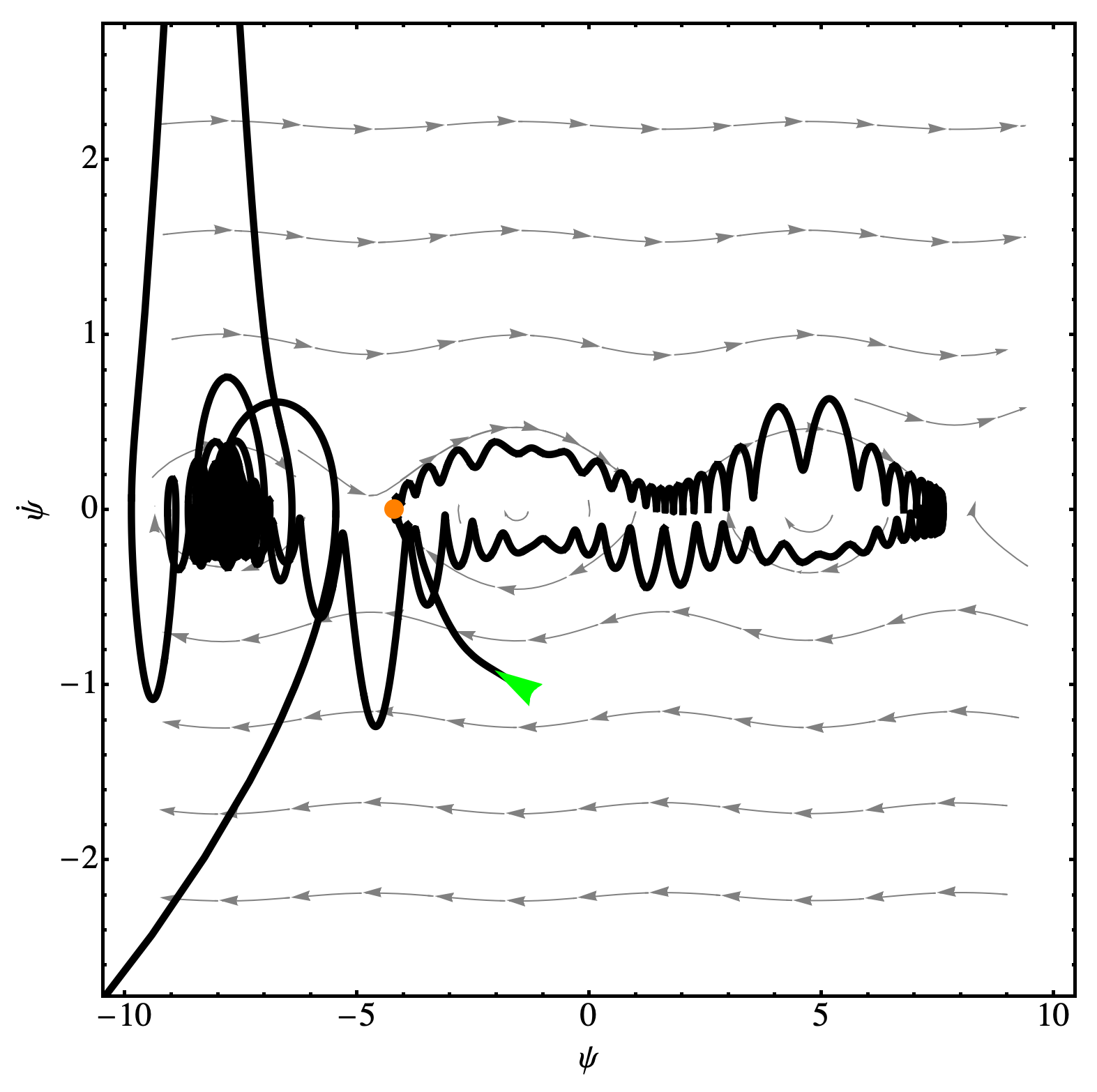}
        \caption{}
    \end{subfigure}
    \caption{Phase portraits of the slow angle $\psi(t)$ (solid black line) obtained by numerical integration of Eq.~\eqref{eq:angles}, superimposed on the vector field of the corresponding pendulum equation~\eqref{eq:numerical} for the different impulse excitations and two damping scenarios. The top panel corresponds to $\lambda_1=5,\lambda_2=0.125$: (a) $v_1=\sqrt{2E_0},v_2=0$ for $E_0=2.0$ (no 1:1 TRC), (b) $v_1=0,v_2=\sqrt{2E_0}$ for $E_0=6.0$ (1:1 TRC); while the bottom panel corresponds to $\lambda_1=0.125,\lambda_2=5$: (c) $v_1=\sqrt{2E_0},v_2=0$ for $E_0=2.0$ (no 1:1 TRC), (d) $v_1=0,v_2=\sqrt{2E_0}$ for $E_0=6.0$ (1:1 TRC). The coupling stiffness is $\varepsilon\beta=0.08$, with $\varepsilon=0.01$, and the nonlinear stiffness is $\alpha=1$. The green arrow indicates the starting point of each trajectory, while the orange dot marks the instant when action $I_1(t)$ reaches the critical level $I_{1,\mathrm{cr}}$.}
    \label{fig:phaseportraits}
\end{figure}

In Fig.~\ref{fig:phaseportraits} we show the realization or absence of 1:1 TRC for different impulse excitations, $v_1=\sqrt{2E_0},v_2=0$ and $v_1=0,\,v_2=\sqrt{2E_0}$. In each plot the green arrow indicates the starting point of each trajectory (which was chosen to be close to the resonance manifold), while the orange dot marks the instant when the action $I_1(t)$ reaches the critical level $I_{1,\mathrm{cr}}$.

In Fig.~\ref{fig:nonlinvelmass2} (a), we have already shown that when the nonlinear DoF is excited, $v_1=\sqrt{2E_0},\,v_2=0$, with $E_0=2.0$, no 1:1 TRC occurs, even though the (necessary) condition~\eqref{eq:cond1} is satisfied. The phase portrait in Figs.~\ref{fig:phaseportraits} (a) clarify that this is due to the absence of an attracting fixed point in the phase space, implying that condition~\eqref{eq:cond2} is not satisfied. In this case, the action of the linear mass $B_0$, at the instant when the action $I_1(t)$ reaches the level $I_{1,\mathrm{cr}}$, is too low for fixed points to exist in the forced pendulum~\eqref{eq:numerical}, so no 1:1 TRC can be realized. In contrast, Fig.~\ref{fig:nonlinvelmass2} (c) shows when the nonlinear DoF is lightly damped and attracting fixed points exist in the phase space.

For the other cases considered -- cf. Figs.~\ref{fig:phaseportraits} (b), (d), however, we observe the existence of fixed points (saddles and centers) which form the 1:1 resonant manifold and may attract the trajectory into capture. The phase space structures are consistent with the 1:1 TRC observed in Fig.~\ref{fig:nonlinvelmass2_2}. The phase portrait Fig.~\ref{fig:phaseportraits} (d) is shown over a wider range of the angle $\psi$ and reveals particularly interesting dynamics, characterized by a long sequence of recurring beating phenomena. This behavior is also visible in the corresponding wavelet transform of the velocity time series shown in Fig.~\ref{fig:nonlinvelmass2_2} (b).

\begin{figure}[ptb]
    \centering
    \begin{subfigure}{0.48\textwidth}
        \centering
        \includegraphics[width=\linewidth]{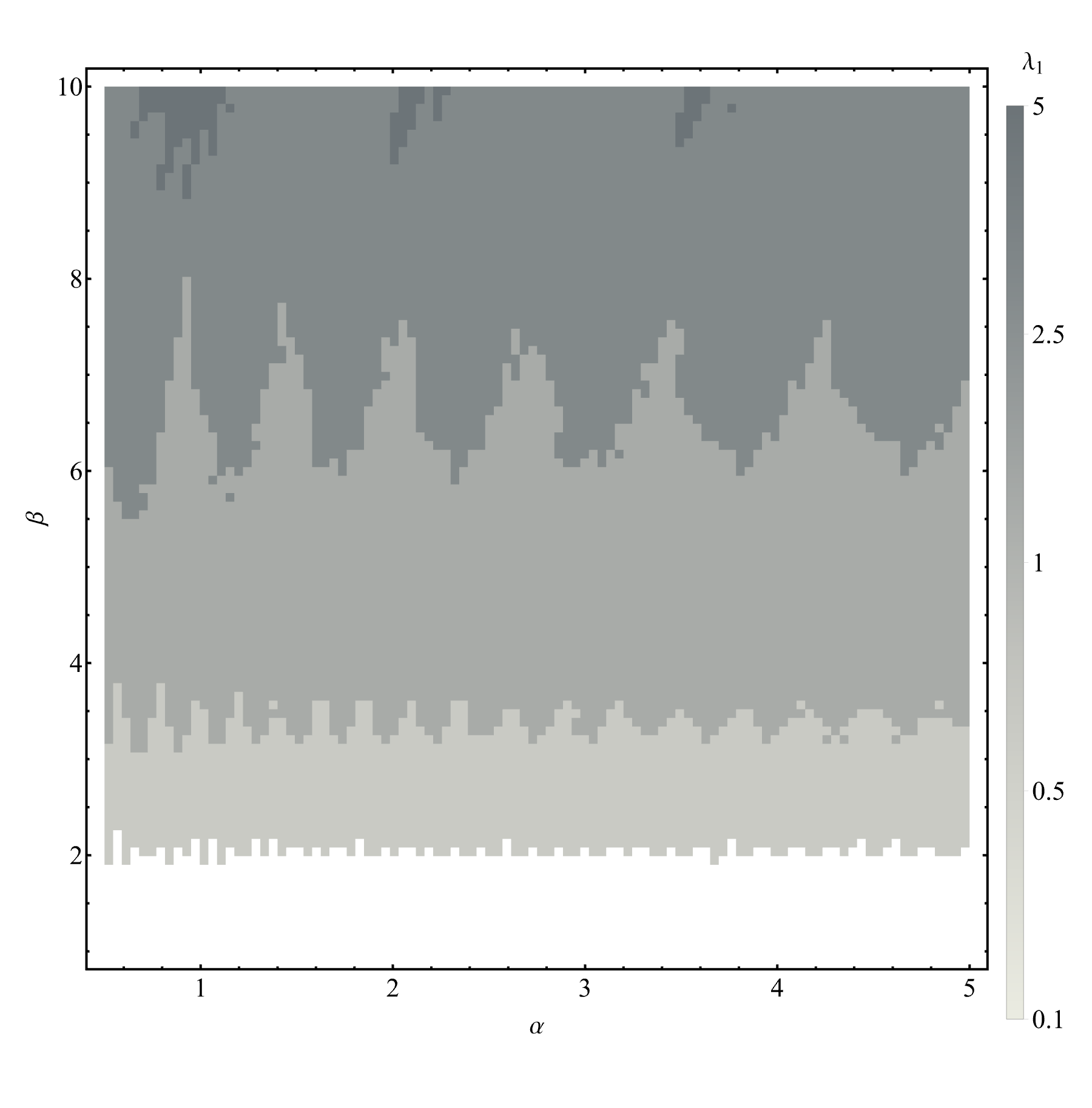}
        \caption{}
    \end{subfigure}
    \hfill
    \begin{subfigure}{0.48\textwidth}
        \centering
        \includegraphics[width=\linewidth]{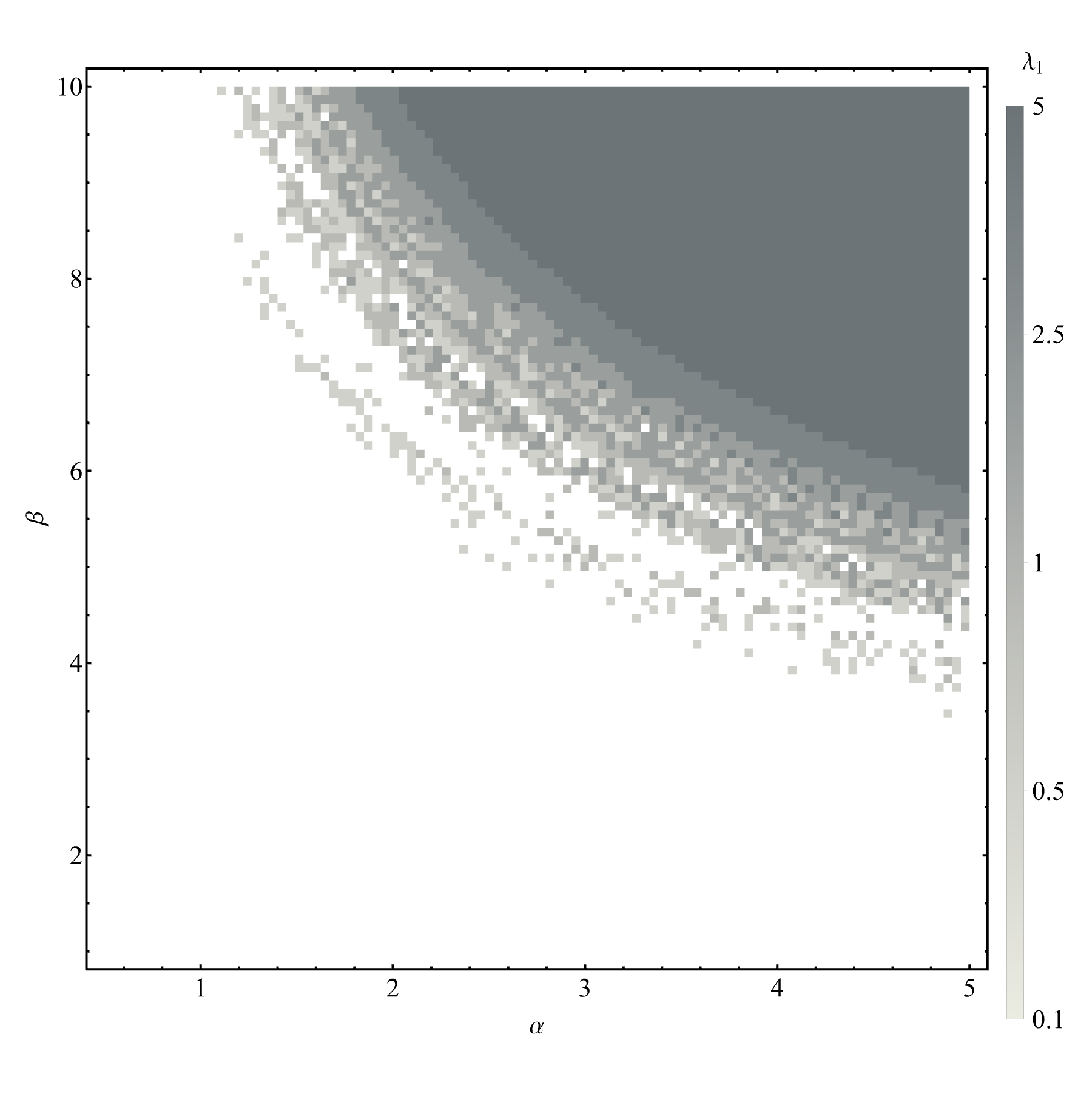}
        \caption{}
    \end{subfigure}

    \vspace{.5cm}

    \centering
    \begin{subfigure}{0.48\textwidth}
        \centering
        \includegraphics[width=\linewidth]{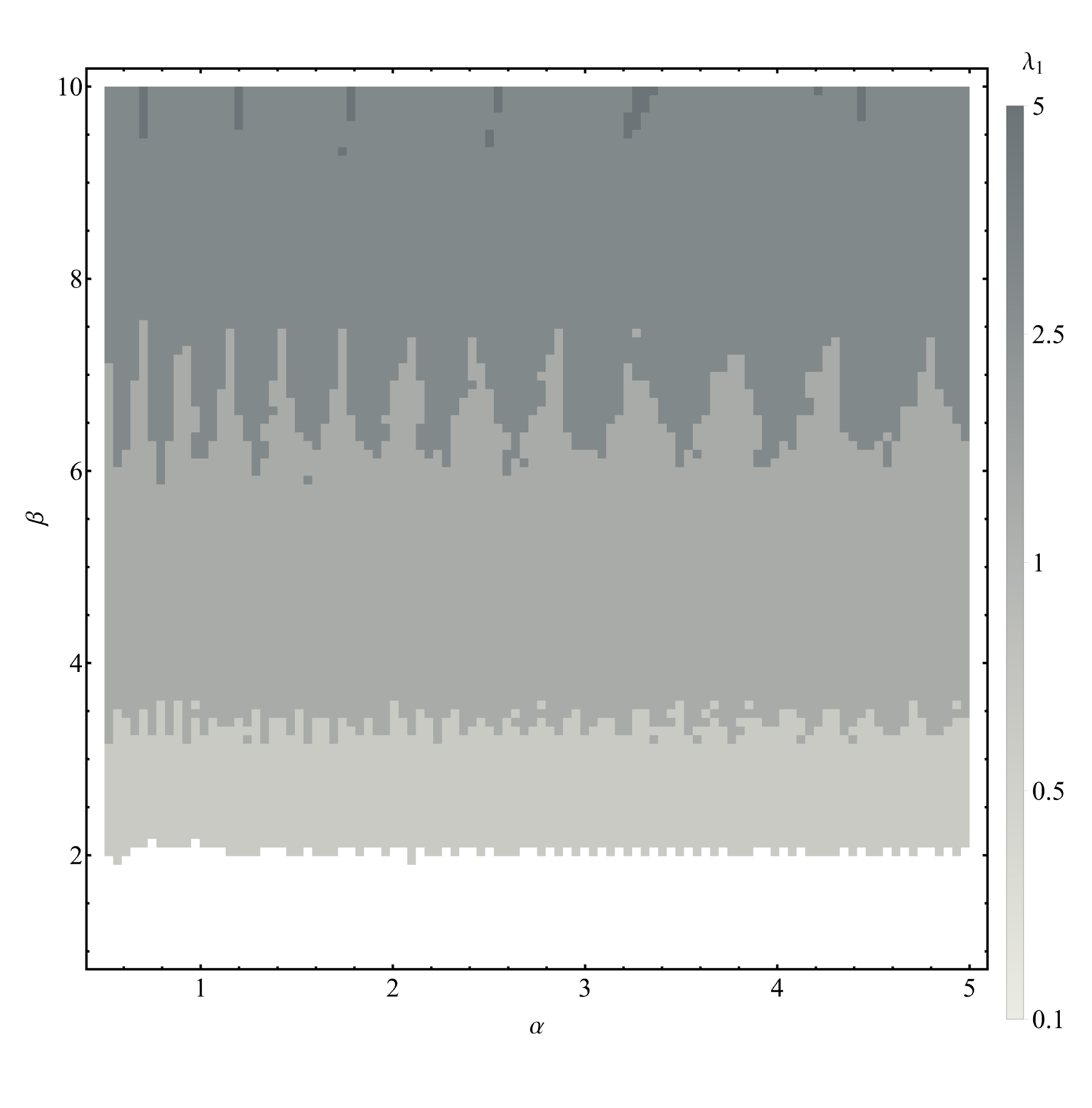}
        \caption{}
    \end{subfigure}
    \hfill
    \begin{subfigure}{0.48\textwidth}
        \centering
        \includegraphics[width=\linewidth]{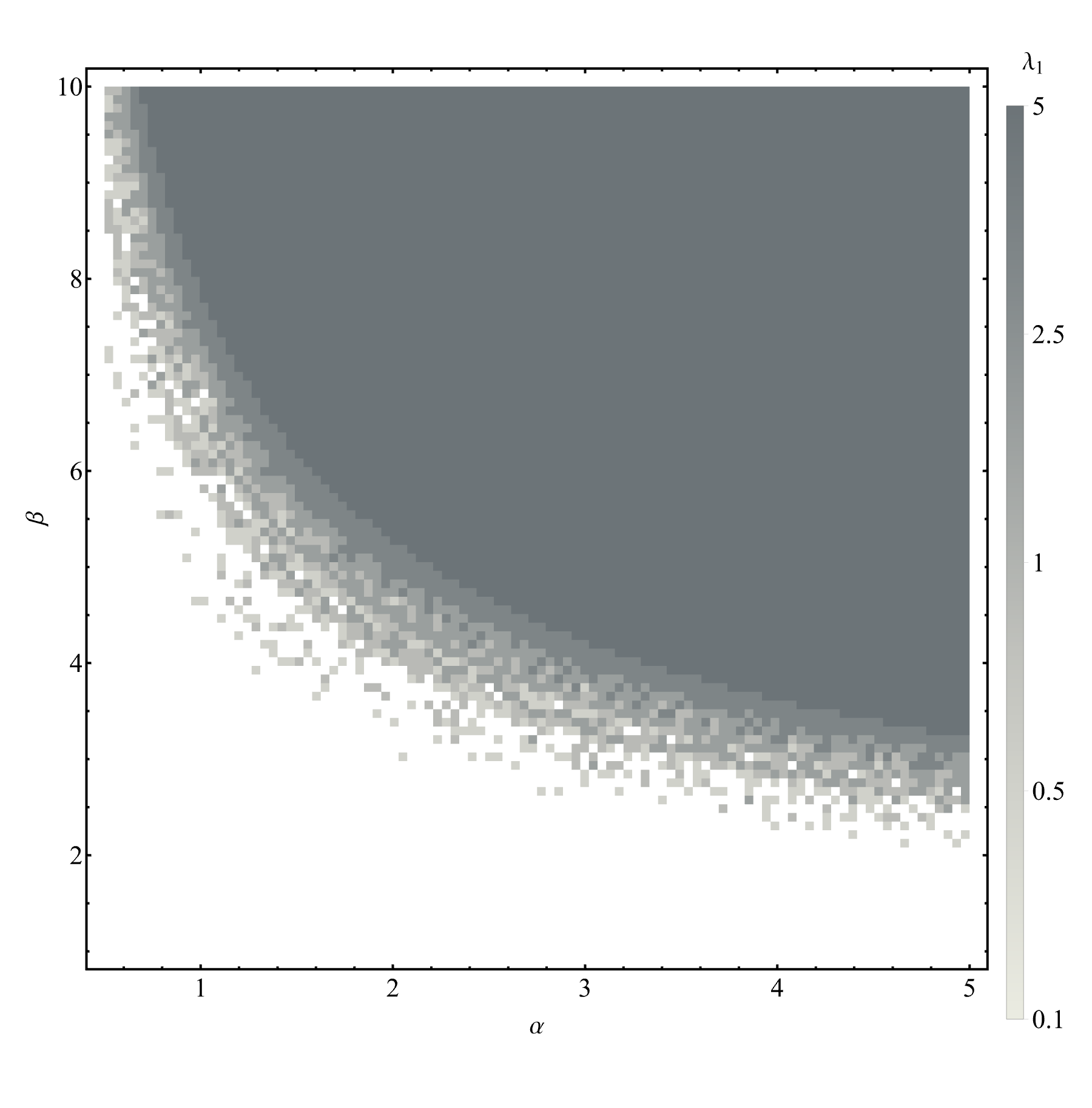}
        \caption{}
    \end{subfigure}
    \caption{Parameter ranges (shaded areas) where the necessary and sufficient conditions for 1:1 TRC are satisfied. Both excitation scenarios are considered: (a), (c) $v_1=\sqrt{2E_0},v_2=0$ and (b), (d) $v_1=0,\,v_2=\sqrt{2E_0}$ are considered, respectively. The remaining parameters are $\varepsilon=0.01$, $\lambda_2=0.125$, with $E_0=2.0$ in (a), (b) and $E_0=6.0$ in (c), (d).}
    \label{fig:boundaries_new}
\end{figure}

In Fig.~\ref{fig:boundaries_new} both impulsive excitation scenarios were considered for $E_0=2.0$ and $E_0=6.0$ and the parameter regions where a 1:1 resonant manifold occurs are depicted for varying damping $\lambda_1$. The regions were obtained by numerically integrating Eq.~\eqref{eq:angles} using the parameter values $\varepsilon=0.01$ and $\lambda_2=0.125$. Event detection was used to determine what the value of $B_0$ is when the action $I_1(t)$ reaches the critical value $I_{1,\mathrm{cr}}$. From Figs.~\ref{fig:boundaries_new} (a), (c) it follows that the damping ratio $\lambda_1$ influences greatly the existence of the resonant manifold only when the nonlinear DoF is excited. In contrast, from Figs.~\ref{fig:boundaries_new} (b), (d) one notes that the effect of $\lambda_1$ on the existence a resonant manifold is small when the linear DoF is excited.

\subsection{Modal interactions and energy exchanges}

To demonstrate explicitly the effect of modal interactions on the decay rates we introduce the super-slow time scale
\begin{equation}
    T_2=\varepsilon \theta_2
\end{equation}
and consider the second-order approximation ($\varepsilon^1$) of $\eta(t)$ as
\begin{equation}
    \begin{aligned}
        \frac{\partial\eta_2}{\partial T_0} &= -\frac{\partial B_0(T_1,T_2)}{\partial T_2}-\frac{\partial B_1(T_1,T_2)}{\partial T_1}
-\Big(2\lambda_2 B_0(T_1,T_2)\cos^2T_0+2\beta B_0(T_1,T_2)\sin T_0\cos T_0
-\\ &-\beta\Lambda I_{1,\mathrm{cr}}^{1/3}\sqrt{2B_0(T_1,T_2)}\,\mathrm{cn}\!\Big(\tfrac{2K(m)}{\pi}(C_0(T_1,T_2)+T_0),m\Big)\cos T_0\Big) \label{eq:secapprox}
    \end{aligned}
\end{equation}

Since 1:1 TRC means that the dynamics in the vicinity of the attracting fixed point satisfies, 
\begin{equation}
    C_{0,\mathrm{e}}^\prime\approx0
\end{equation}  
and from Eq.~\eqref{eq:pend},
\begin{equation}
    A_0(T_1)\approx0 \label{eq:Aconst}
\end{equation}
The action (and consequently the energy) of the nonlinear DoF is nearly conserved on the resonant manifold, i.e., it is an adiabatic invariant at the fast time scale \cite{Verhulst1996}. This means, that the average over one $2\pi$-period in $T_0$ of the right-hand side of
Eq.~\eqref{eq:secapprox} carries no $T_1$ dependence once the phase is locked near the resonant manifold, implying, 
\begin{equation}
    \frac{\partial B_1(T_1,T_2)}{\partial T_1}\equiv0
\end{equation}
and ultimately that,
\begin{equation}
\frac{\partial B_0(T_2)}{\partial{T_2}} = -\lambda_2 B_0(T_2) + c_\beta\beta\Lambda I_{1,\mathrm{cr}}^{1/3}\sqrt{2B_0(T_2)}\cos C_0(T_2) \label{eq:B0prime}
\end{equation}
where $c_\beta=0.4775$ represents a constant from the Fourier projection of the elliptic term on the cosine term in the averaging with respect to the fast time scale $T_0$:
\begin{equation}
    \left\langle\mathrm{cn}\left(\frac{2K(m)}{\pi}(C_0(T_2)+T_0),m\right)\cos\theta_2(t)\right\rangle=c_\beta\cos C_0(T_2)
\end{equation}
With Eq.~\eqref{eq:Aconst}, Eq.~\eqref{eq:pend2} reduces to,
\begin{equation}
    \cos C_0(T_2)=-\frac{4K^4(m)(1+e^{-\pi})}{3\pi^4 e^{-\pi/2}}\frac{\lambda_1}{\beta\sqrt{\alpha B_0(T_2)}}\label{eq:pend3}
\end{equation}
which, when substituted into Eq.~\eqref{eq:B0prime} yields,
\begin{equation}
\frac{\partial B_0(T_2)}{\partial{T_2}}=-\lambda_2B_0(T_2)-\frac{8\sqrt{2}c_\beta K^5(m)(1+e^{-\pi})}{3\pi^5 e^{-\pi/2}}\frac{\lambda_1}{\alpha} \label{eq:B2}
\end{equation}
or,
\begin{equation}
\frac{1}{B_0(T_2)}\frac{\partial B_0(T_2)}{\partial T_2}=-\left(\lambda_2+0.6470\frac{\lambda_1}{\alpha B_0(T_2)}\right) \label{eq:B0primenorm}
\end{equation}
revealing the combination of both damping ratios $\lambda_1$ and $\lambda_2$ in the modal interactions of the nonlinear system~\eqref{eq:nonlineom}; this differs from the linear case where only the average of the two damping ratios $\lambda_s$, appears.

Equation~\eqref{eq:B2} is a linear ODE with the solution (after restoring $T_2=\varepsilon t$):
\begin{equation}
    B_0(\varepsilon t) = \left(B_0(0) + \frac{0.6470}{\alpha}\frac{\lambda_1}{\lambda_2}\right)e^{-\varepsilon\lambda_2 t} - \frac{0.6470}{\alpha}\frac{\lambda_1}{\lambda_2} \label{eq:B0sol}
\end{equation}
This gives the approximate slow decay of the linear-DoF action, and consequently of the total energy $E(\varepsilon t)\approx\frac{\alpha\Lambda^4}{4}I^{4/3}_{1,\mathrm{cr}}+ B_0( \varepsilon t)$ during 1:1 TRC.

\begin{figure}[ptb]
    \centering
    \begin{subfigure}{0.48\textwidth}
        \centering
        \includegraphics[width=\linewidth]{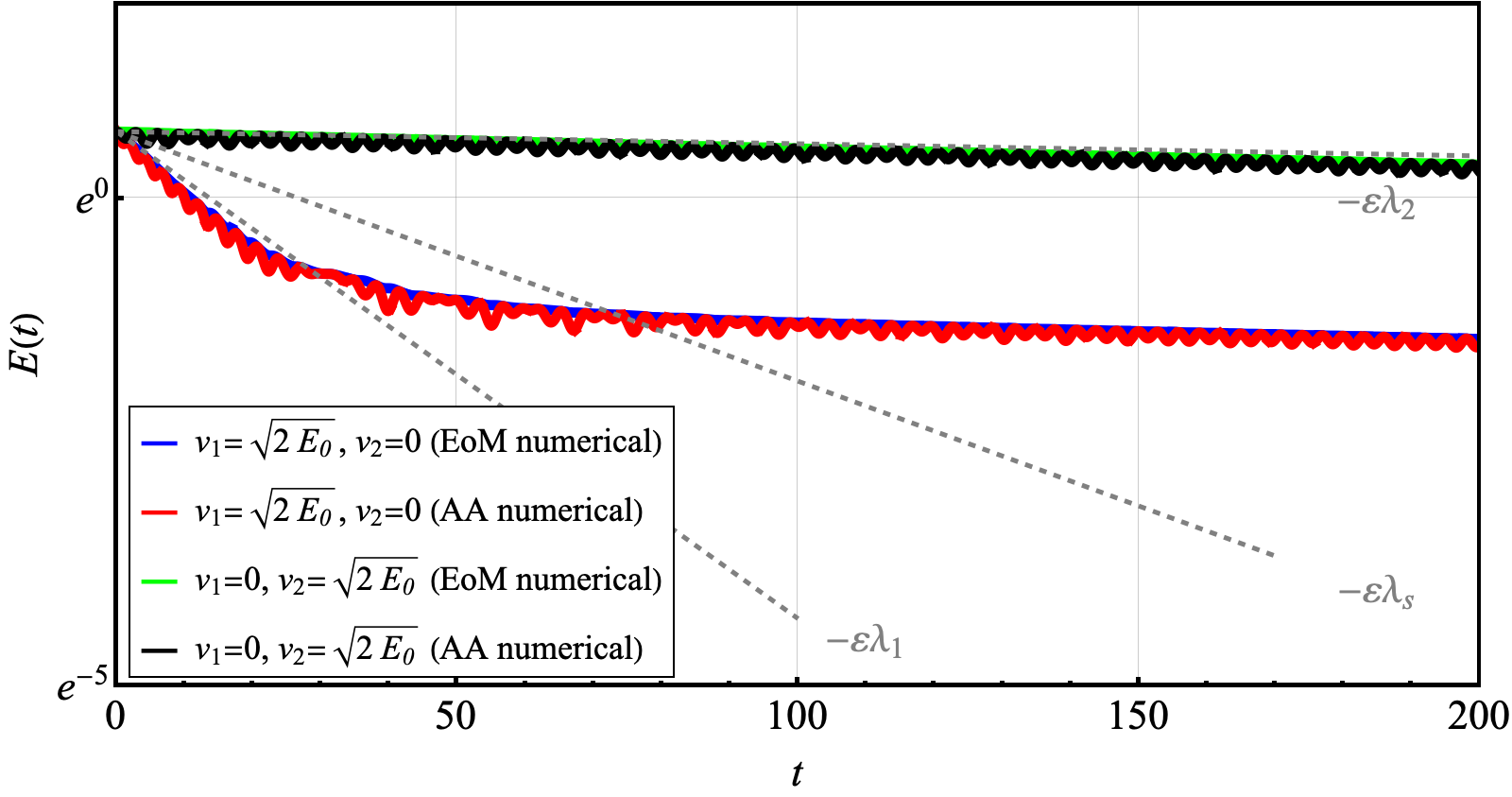}
        \caption{}
    \end{subfigure}
    \hfill
    \begin{subfigure}{0.48\textwidth}
        \centering
        \includegraphics[width=\linewidth]{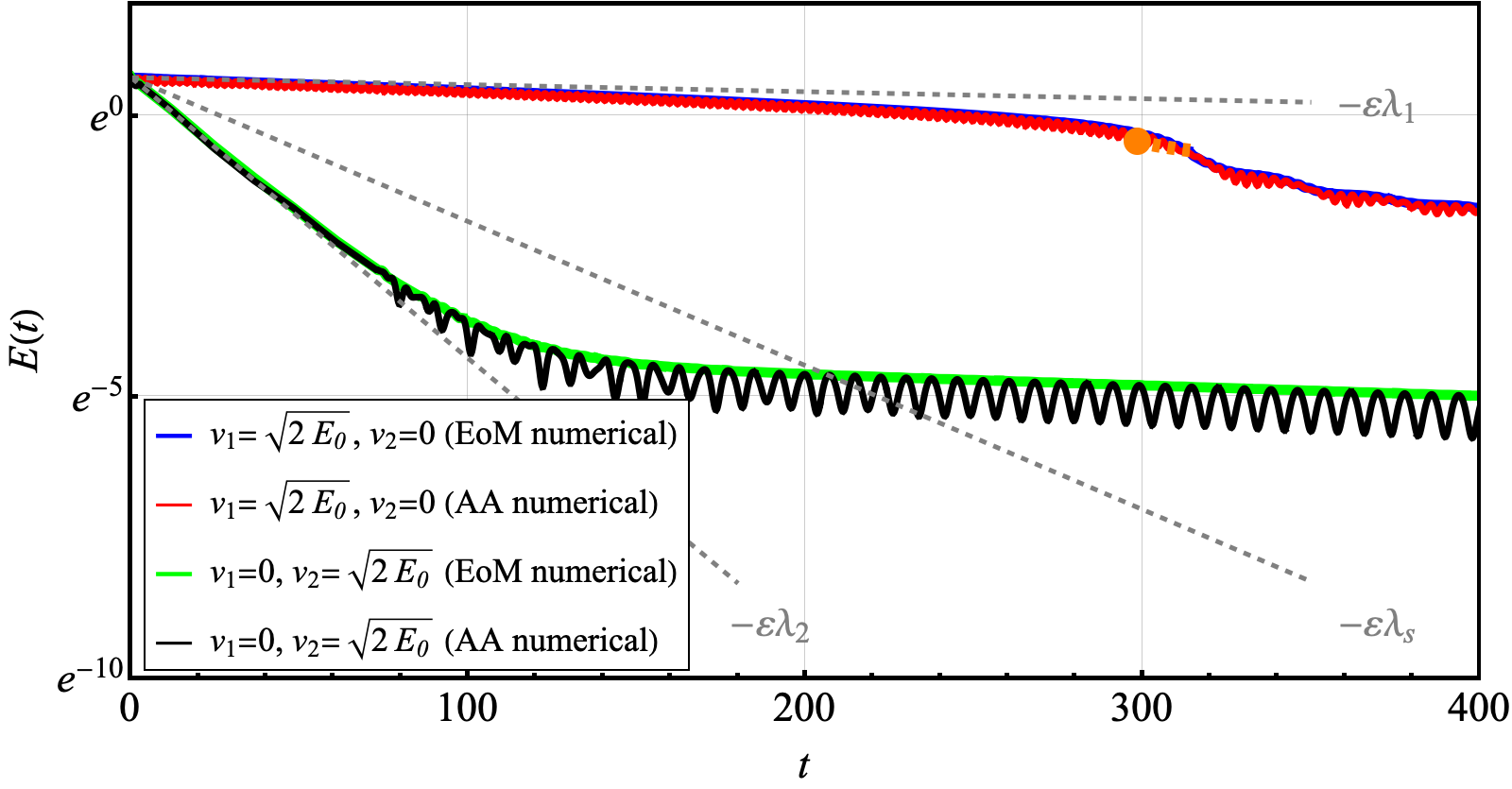}
        \caption{}
    \end{subfigure}

    \vspace{0.5cm}

    \begin{subfigure}{0.48\textwidth}
        \centering
        \includegraphics[width=\linewidth]{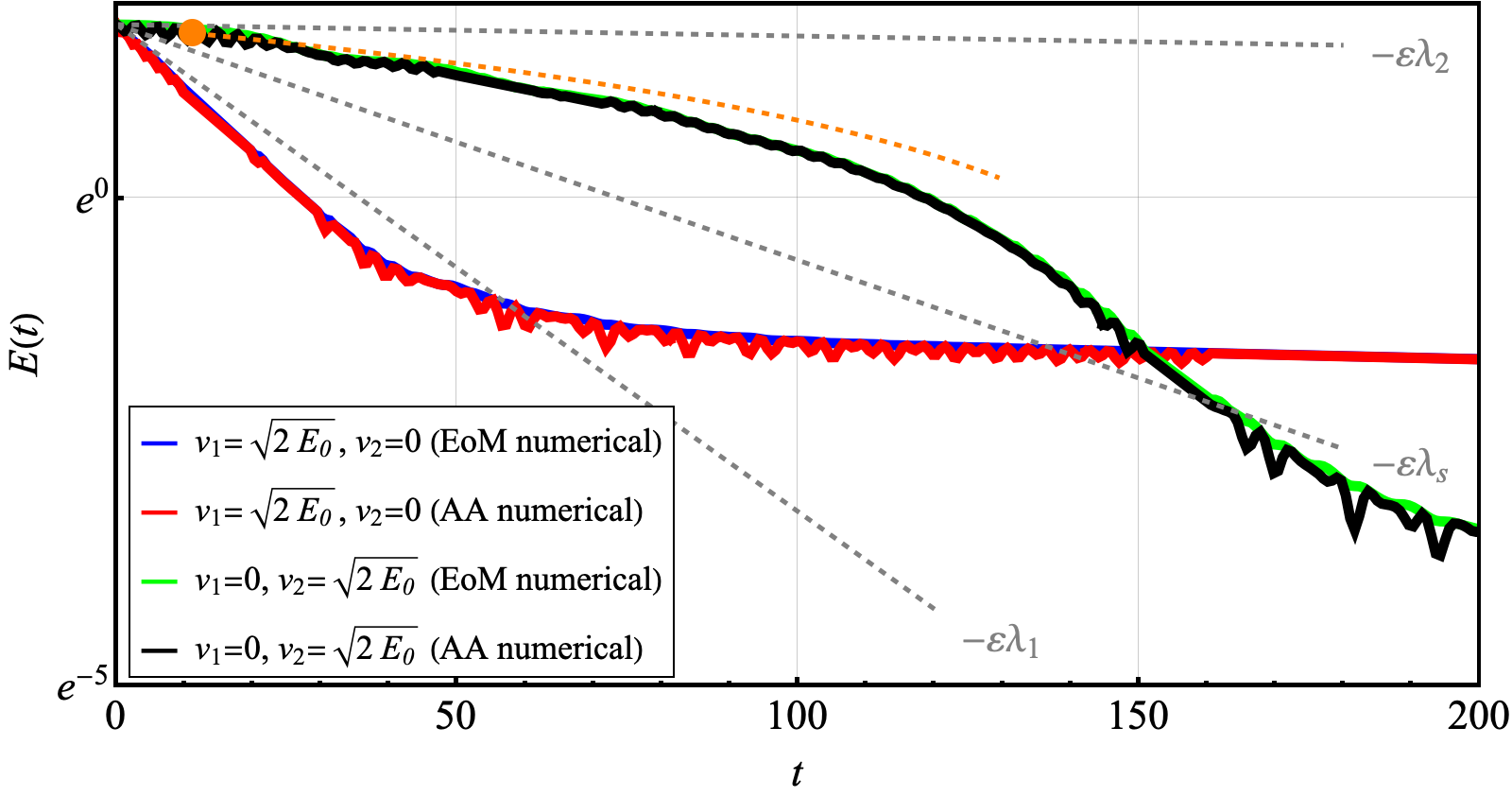}
        \caption{}
    \end{subfigure}
    \hfill
    \begin{subfigure}{0.48\textwidth}
        \centering
        \includegraphics[width=\linewidth]{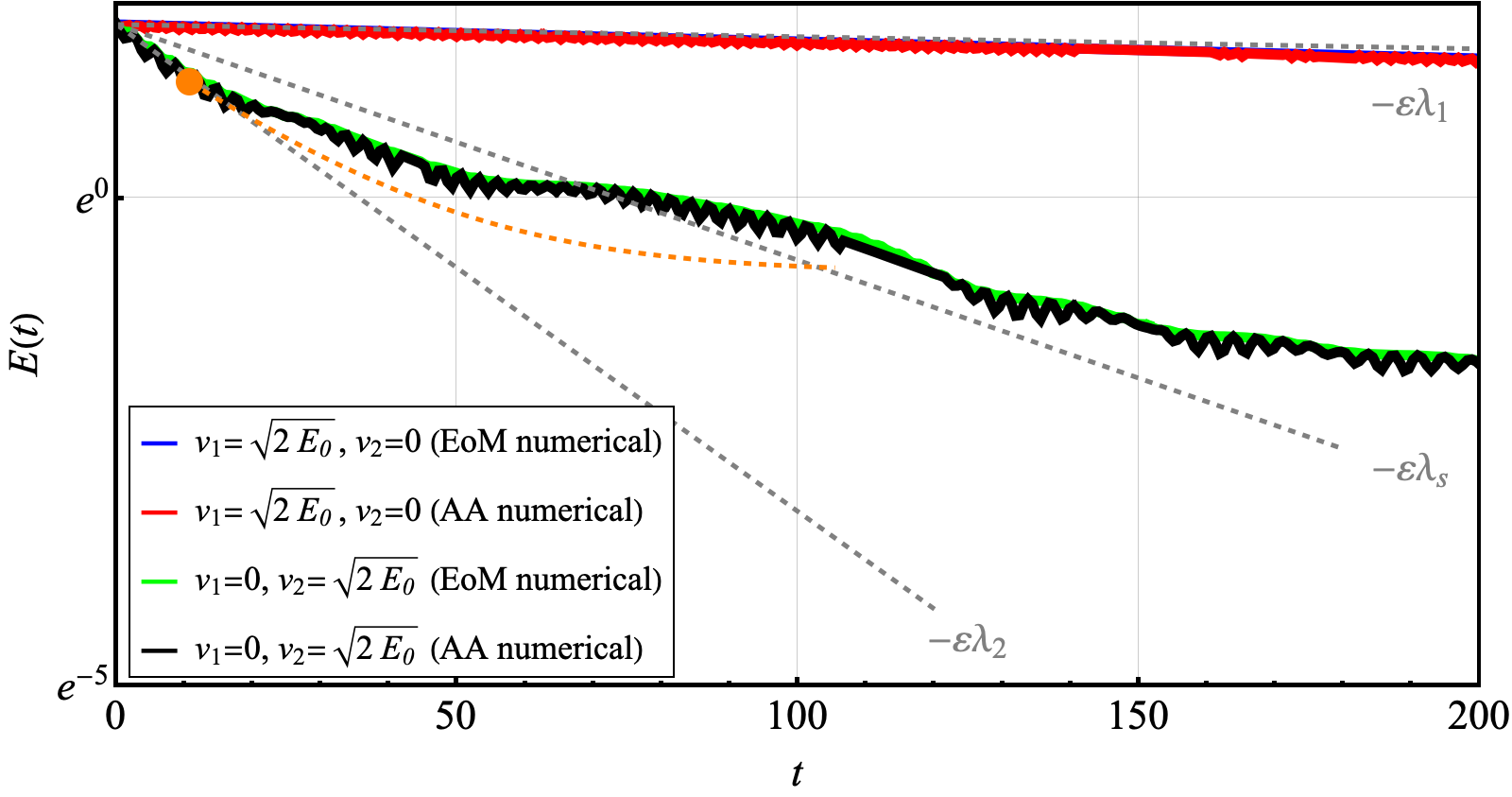}
        \caption{}
    \end{subfigure}
    \caption{Comparison of the logarithmic energy $E(t)$ of the 2 DoF nonlinear coupled oscillator~\eqref{eq:nonlineom} for two impulse excitations, $v_1=\sqrt{2E_0},v_2=0$ and $v_1=0,v_2=\sqrt{2E_0}$, and two initial energies: (a), (b) $E_0=2.0$ (no TRC) and (c), (d) $E_0=6.0$ (TRC). The coupling stiffness is $\varepsilon\beta=0.08$, with $\varepsilon=0.01$, and the nonlinear stiffness is $\alpha=1$. Panels (a), (c) correspond to $\lambda_1=5,\lambda_2=0.125$ and panels (b), (d) to $\lambda_1=0.125,\lambda_2=5$. The solid blue and green lines are obtained by numerical integration of system~\eqref{eq:nonlineom}, whereas the red and black lines represents the approximations of the energy by \eqref{eq:angles}. The orange dot denotes when $I_1(t)$ reaches the $I_{1,\mathrm{cr}}$ level and the orange dashed line represents the approximation of energy decay in 1:1 TRC based on Eq.~\eqref{eq:B0sol}. The orange dashed line is plotted until $0.9I_{1,\mathrm{cr}}$ is reached where Eq.~\eqref{eq:B0sol} is not valid anymore.
    }
    \label{fig:nonlinenergies}
\end{figure}

Fig.~\ref{fig:nonlinenergies} shows the total energy decay $E(t)$ of system~\eqref{eq:nonlineom}, on a logarithmic scale, for the two excitation cases $v_1=\sqrt{2E_0},\,v_2=0$ and $v_1=0,\,v_2=\sqrt{2E_0}$, and for two initial energies: $E_0=2.0$ in panels (a), (b) and $E_0=6.0$ in panels (c), (d). The coupling stiffness is $\varepsilon\beta=0.08$ with $\varepsilon=0.01$, and the nonlinear stiffness is $\alpha=1$. Panels (a), (c) correspond to $\lambda_1=5,\,\lambda_2=0.125$, and panels (b), (d) to $\lambda_1=0.125,\,\lambda_2=5$. 

Figs.~\ref{fig:nonlinenergies} (a), (c) ($\lambda_1>\lambda_2$) further reinforce the observations from the time series in Figs.~\ref{fig:nonlinvelmass2} and~\ref{fig:nonlinvelmass2_2}. Regardless of the initial energy, when the nonlinear DoF is heavily damped and the initial impulse is applied to this DoF, no 1:1 TRC, and consequently, no significant modal interactions can occur. In this case, the energy of system~\eqref{eq:nonlineom} is dissipated primarily through the nonlinear DoF, which has a damping ratio of $\varepsilon\lambda_1$. As the oscillation amplitude decreases sufficiently, the remaining energy is dissipated through the linear DoF with a damping ratio of $\varepsilon\lambda_2$. In contrast, when a sufficiently large initial impulse is applied to the linear DoF Fig.~\ref{fig:nonlinenergies} (c), 1:1 TRC occurs, resulting in modal interactions, and the system is damped at an effective damping ratio close to $\varepsilon\lambda_s$; but rather it is governed by Eq.~\eqref{eq:B0sol} (denoted by orange dashed line) during the time of the 1:1 TRC.

Fig.~\ref{fig:nonlinenergies} (b) exhibits similar behavior to the case of $E_0=2.0$, however, although the conditions for resonance capture are not fulfilled for excitation of the linear DoF, the case of the nonlinear DoF excited may still be of interest. Therefore, we show the energy evolution over a longer time interval. As observed in Fig.~\ref{fig:nonlinenergies} (b), when $\lambda_1<\lambda_2$, the onset of modal interactions takes longer time. Fig.~\ref{fig:nonlinenergies} (b) shows that energy is eventually transferred to the heavily damped linear DoF. Although this transfer occurs only over a relatively short time interval, the energy evolution approaches a slope that is combination of $\lambda_1$ and $\lambda_2$ given by~\eqref{eq:B0sol} (orange dashed line). This emphasizes that in the nonlinear system the energy decay through modal interaction is not described by $\lambda_s$ as predicted by linear theory.

For $E_0=6.0$ and $\lambda_1<\lambda_2$, shown in Fig.~\ref{fig:nonlinenergies} (d), modal interactions through 1:1 TRC occur for both excitation cases. Consequently, during the 1:1 TRC the energy decreases with a slope corresponding to Eq.~\eqref{eq:B0sol} (orange dashed line). However, here we can clearly observe this phenomenon only in the excitation case $v_1=0,v_2=\sqrt{2E_0}$. Since $\lambda_1$, the damping coefficient of the nonlinear DoF, is small it takes a longer time to reach the moment of resonance capture, as was demonstrated previously by Fig.~\ref{fig:nonlinenergies} (b).

We refer the reader to the \hyperref[appendix]{Appendix} where there is four supplementary animations showcasing the total energy decay of system~\eqref{eq:nonlineom} (along the corresponding phase locking mechanisms as shown in Fig.~\ref{fig:phaseportraits}) for different values of initial energy at given nonlinear stiffness, coupling and damping ratios.

\section{Concluding remarks}
\label{sec:concl}

In this work, we analyzed the modal interactions in a 2 DoF linear, as well as a nonlinear coupled oscillator and closely spaced modes with non-classical viscous damping distribution. The analysis was carried out by utilizing the action-angle formulation of the systems and leading order approximations were obtained by the method of multiple scales.

For the linear case we recovered the results already existing in the literature. Namely, that the non-classical damping distribution couples the symmetric and antisymmetric modes, thus, resulting in the two regimes of different transient dynamics.  

The strongly nonlinear case is drastically different, however. The absence of linear stiffness, together with the nonlinear cubic stiffness, couples the symmetric and antisymmetric modal coordinates independently of the damping distribution; thus, modal interactions can occur even under proportional damping. The magnitudes of the damping ratios govern the existence of the 1:1 resonant manifold as well as the energy decay characteristics of the response, specifically, the extent to which the effective decay rate deviates from the individual damping rates. We identify the necessary and sufficient conditions (and the corresponding ranges) for which the action-dependent frequency of the nonlinear DoF can reach the 1:1 resonance manifold and for the transient dynamics of the system to be captured in 1:1 TRC. To this end, we derive the governing slow-flow equation for the slow angle, which takes the form of a forced pendulum equation. We then obtain an analytical approximation for the energy-decay slope during 1:1 TRC. Together, these results provide analytical, rather than purely numerical, access to the parameter regions in which strongly nonlinear modal interactions occur and the mechanism through which resonance capture couples the two dissipation rates. Our analytical results conform with direct numerical simulations.

We foresee interesting potential future applications of this methodology. The problem of modal interactions in structures with closely spaced modes due to mechanical joints is one possible application; in that case, the nonlinear stiffness term of the present work could be replaced by a friction-hysteresis model which not only will break the symmetry in the damping distribution, but also introduce nonlinear effects. Another interesting application concerns the extension of the current results to cyclic structures (e.g., bladed disk arrays) where it would be of interest to analyze modal interactions due to non-proportional damping distribution caused by mistuning, and nonlinear modal interactions due to friction, aeroelastic coupling and other effects. Hence, the present work paves the way to rigorous study of a diverse class of problems that are of practical interest to multi-physics and multi-scale mechanical design.

\section*{Acknowledgements}

Csanád Á. Hubay gratefully acknowledges financial support for this research project from the Hun\-garian-American Fulbright Program, sponsored by the U.S. Department of State and the Fulbright Commission, Hungary. The work of Csanád Á. Hubay is also supported by the Doctoral Excellence
Fellowship Program (DCEP-25-1-BME-54), funded by the
National Research Development and Innovation Fund of the Ministry of Culture and Innovation and the Budapest University of Technology
and Economics.

\section*{Appendix}

\subsection*{Supplementary derivation and animations for the nonlinear system}
\label{appendix}
The derivation of the action-angle formulation~\eqref{eq:actionangles} of the nonlinear DoF is provided in a
\href{https://github.com/csan94/2DoF_nonlinear_oscillator/blob/main/supplementary_material.pdf}{supplementary document}. 

To complement Figs.~\ref{fig:phaseportraits} and~\ref{fig:nonlinenergies}, we provide as \href{https://github.com/csan94/2DoF_nonlinear_oscillator/tree/main}{online supplementary material} four animations. In each, $\varepsilon = 0.01$ and the indicated mass is impulsively forced with integer initial energies $E_0 \in [1.0, 8.0]$. The red dot in the left panel marks the simulated parameter set in the parameter space; the middle panel shows the phase portraits, and the right panel the corresponding total energy decay.

\subsubsection*{Animation 1}

Mass $m_1$ is forced, with $\alpha = 0.5$, $\beta = 10$, $\lambda_1 = 0.125$, and
$\lambda_2 = 2.5$. 1:1 TRC occurs for all $E_0$ except $E_0 = 1.0$; the higher $E_0$,
the later $I_1(t)$ reaches $I_{1,cr}$. The capture is short-lived: the energy
transferred to the heavily damped linear DoF is dissipated rapidly, which appears as a
sudden drop in $E(t)$.

\subsubsection*{Animation 2}

Mass $m_1$ is forced, with $\alpha = 1$, $\beta = 8$, $\lambda_1 = 5$, and
$\lambda_2 = 0.125$. For every $E_0$ the parameter set lies outside the TRC region: when
$I_1(t)$ reaches $I_{1,cr}$, the action of the linear DoF is too small for an attracting
fixed point to exist, so $\psi-\dot{\psi}$ trajectory passes through resonance without capture and the energy
decays first at $\varepsilon\lambda_1$, then at $\varepsilon\lambda_2$
(cf. Figs.~\ref{fig:phaseportraits} (a) and \ref{fig:nonlinenergies} (a), (c)).

\subsubsection*{Animation 3}

Mass $m_2$ is forced, with $\alpha = 1$, $\beta = 8$, $\lambda_1 = 2.5$, and
$\lambda_2 = 0.125$. 1:1 TRC occurs only for $E_0 \geq 6.0$; for lower $E_0$ the energy
transferred through the coupling cannot drive $I_1(t)$ to $I_{1,cr}$. During capture,
$I_1(t)$ is held near $I_{1,cr}$ while the linear DoF feeds energy to the heavily damped
nonlinear DoF, so the energy decays at the combined rate of Eq.~\eqref{eq:B0sol}.

\subsubsection*{Animation 4}

Mass $m_2$ is forced, with $\alpha = 2$, $\beta = 6$, $\lambda_1 = 0.125$, and
$\lambda_2 = 5$. The mechanism is the same as in Animation~3, but with the linear DoF
heavily damped, the modal interactions cause a plateau in the energy decay, and the
total energy decay is not monotonic with respect to $E_0$.

\subsection*{Declaration of generative AI use}
The supplementary animations were prepared with the assistance of Claude (Anthropic). Based on the mechanical model, action-angle formulation, slow-flow analysis and resonance-capture conditions derived by the authors (Eqs.~\eqref{eq:nonlineom}, \eqref{eq:angles}, \eqref{eq:cond}, \eqref{eq:numerical}, \eqref{eq:cond2}, and \eqref{eq:B0sol}), Claude wrote and executed the Python code that integrates the equations of motion and action-angle equations, computes the parameter regions, and renders the frames. The authors specified the cases and parameters, checked the results against their own independent simulations, and take full responsibility for the animations.

\bibliographystyle{authordate1}
\bibliography{ref}

 \end{document}